\documentclass[twocolumn]{aastex631}
\usepackage{fontawesome}
\usepackage[T1]{fontenc}
\usepackage{times}

\accepted{September 29, 2025}

\begin{document}

\title{A Compact Symmetric Object Discovered by the VLA Low-band Ionosphere and Transient Experiment}

\author[0000-0003-1991-370X]{Kristina Nyland}
\altaffiliation{Present affiliations: Computational Physics Inc, 8001 Braddock Rd \# 210, Springfield, VA 22151, USA; U.S. Naval Observatory, 3450 Massachusetts Ave NW, Washington, DC 20392}
\affiliation{U.S. Naval Research Laboratory, 4555 Overlook Ave SW, Washington, DC 20375, USA}

\author{Mary Rachelle Barrett}
\affiliation{U.S. Naval Academy, 290 Buchanan Rd, Annapolis, MD 21402, USA}

\author[0009-0003-4877-948X]{Genna Crom}
\affiliation{Physics Department, New Mexico Tech, Socorro, NM 87801, USA}

\author[0000-0002-9471-8499]{Pallavi Patil}
\affiliation{William H. Miller III Department of Physics and Astronomy, Johns Hopkins University, Baltimore, MD 21218, USA}

\author[0000-0003-3272-9237]{Emil Polisensky}
\affiliation{U.S. Naval Research Laboratory, 4555 Overlook Ave SW, Washington, DC 20375, USA}

\author[0000-0003-3272-9237]{Wendy Peters}
\affiliation{U.S. Naval Research Laboratory, 4555 Overlook Ave SW, Washington, DC 20375, USA}

\author[0000-0003-3272-9237]{Simona Giacintucci}
\affiliation{U.S. Naval Research Laboratory, 4555 Overlook Ave SW, Washington, DC 20375, USA}

\author[0000-0001-6812-7938]{Tracy Clarke}
\affiliation{U.S. Naval Research Laboratory, 4555 Overlook Ave SW, Washington, DC 20375, USA}

\author[0000-0002-3032-1783]{Mark Lacy}
\affiliation{National Radio Astronomy Observatory, Charlottesville, VA 22903, USA}

\author{Shyaam Mukundan}
\altaffiliation{Present affiliation: Department of Computer Science, University of Maryland, College Park, MD, 20742}
\affiliation{Heritage High School, 520 Evergreen Mills Rd SE, Leesburg, VA 20175, USA}

\author[0000-0001-9584-2531]{Dillon Z. Dong}
\affiliation{National Radio Astronomy Observatory, P.O. Box 0, Socorro, NM 87801, USA}

\author[0000-0003-4700-663X]{Andy Goulding}
\affiliation{Department of Astrophysical Sciences, Princeton University, Princeton, NJ 08544, USA}

\author[0000-0001-9324-6787]{Amy E Kimball}
\affiliation{National Radio Astronomy Observatory, 1011 Lopezville Rd., Socorro, NM 87801, USA}

\author[0000-0002-6741-9856]{Magdalena Kunert-Bajraszewska}
\affiliation{Institute of Astronomy, Faculty of Physics, Astronomy and Informatics, NCU, Grudziądzka 5/7, 87-100, Toruń, Poland}




\begin{abstract}
We present new Very Long Baseline Array (VLBA) imaging of a MHz-peaked spectrum (MPS) source that was found using commensal low-frequency data taken with the Karl G. Jansky Very Large Array (VLA). The source, J0330-2730, was identified in multi-epoch data from the VLA Low-band Ionosphere and Transient Experiment (VLITE).  VLITE continuously collects low-frequency data at 340~MHz during regular VLA observations.  Our analysis of the VLITE light curve demonstrates that J0330-2730 has significant 340~MHz flux variability at the $\sim$20\% level over a timescale of approximately one year. Our VLBA images reveal a resolved, double-lobed morphology with a projected linear size of 64~pc.  We consider plausible mechanisms that could explain the observed 340~MHz variability and the source properties on milliarcsecond scales.  We rule-out variable Doppler boosting and conclude that refractive interstellar scintillation or variable free-free absorption are the most likely explanations. We argue that the properties of J0330-2730 are consistent with the class of compact symmetric objects (CSOs) and consider the evolutionary stage of the source.  The extent of the resolved lobes revealed by the VLBA is significantly smaller than predictions based on the turnover-size relation for a standard synchrotron self-absorbed jet model.  We discuss possible explanations for the departure from the turnover-size relation, including jet formation by a transient phenomenon such as a tidal disruption event or a ``frustrated jet'' impeded by the presence of dense gas or a high-pressure environment.   This study highlights the potential of VLITE for the identification of compact and young radio sources. 
\end{abstract}


\section{Introduction} 
\label{sec:intro}
The centers of most massive galaxies harbor a supermassive black hole (SMBH) weighing millions to billions of solar masses (e.g. \citealt{kormendy+95}).  Although SMBHs do not emit any light directly, electromagnetic radiation is emitted during the process of SMBH accretion.  An actively feeding SMBH, known as an active galactic nucleus (AGN) or quasar, may shine thousands of times brighter than an entire galaxy \citep{antonucci+93, urry+95, heckman+14, netzer+15}.  
A subset of all quasars ($\sim1-10$\%) are detected at radio frequencies. 
\citep{best+05, ivezic+02}.  
The radio emission detected in these quasars originates from bipolar jets/lobes of plasma produced by synchrotron-emitting electrons accelerated to relativistic speeds by strong magnetic fields near the SMBH (\citealt{blandford+19}, and references therein).  

Over millions of years, radio jets may grow to large scales (10's to 1000's of kpc), extending far beyond the stellar extents of their host galaxies.  
The energy released by large-scale quasar jets and lobes is believed to generate energetic feedback that regulates the rate and efficiency of galactic star formation (e.g. \citealt{morganti+17, hardcastle+20}).  This process is well understood for large-scale radio galaxies capable of heating reservoirs of gas in the intracluster medium (ICM).  Gas heated by large-scale radio lobes forms bubbles that inhibit the formation of cooling flows, which would otherwise condense and lead to the formation of new stars \citep{fabian+12}.  This mode of feedback is prevalent among massive galaxies residing in dense environments at low redshifts.  
However, much less is known about the physics and impact of jet-driven feedback on sub-galactic scales, where jets in the early stages of their evolution may interact with the interstellar medium (ISM) and influence galaxy properties (e.g. \citealt{kukreti+24}, and references therein).

\subsection{Identifying Young Jets}
Systematic studies of jets that are young ($\lesssim$ 10$^4$ years old) and compact ($\lesssim$ 1~kpc), as well as the properties of the galaxies in which they reside, are thus of fundamental importance for improving our understanding of the life cycles of radio AGN and their connection to galaxy evolution. 
The first step is identifying radio AGN that are young.  Young radio AGN may be identified on the basis of their radio morphologies, spectral shapes, or variability (e.g. \citealt{odea+21}).  

The most direct way to 
identify candidate young/compact jets is by spatially resolving their morphologies on kpc or pc scales with high-angular-resolution imaging (e.g. \citealt{peck+00, tremblay+09}). 
However, the high observing cost of these observations poses logistical challenges.  For instance, a 10~pc-scale radio jet at $z=1$ has an angular size of 1 milliarcsecond.  Observations utilizing the technique of very long baseline interferometry (VLBI) are necessary to achieve such a high angular resolution.  
VLBI is a powerful technique for direct studies of radio AGN evolution.  In some cases, VLBI observations of resolved young jets taken over multiple epochs may be used to directly measure the expansion of the jets and determine their speed (e.g. \citealt{odea+21}, and references therein).  

Historically, blind imaging searches with milliarcsecond-scale resolution have not been possible, and the systematic identification of young radio AGN based on directly resolved morphologies has been limited \citep{peck+00, sokolovsky+11, tremblay+16}.  Thus, this technique has been best suited for confirming the morphologies of candidate young radio AGN identified using different strategies better suited to unbiased searches.   
However, we note that recent advancements with the VLBI capabilities of telescopes like the Low Frequency Array (LOFAR; \citealt{van_haarlem+13}) are enabling wide-field low-frequency radio imaging at subarcsecond resolution \citep{morabito+25}. 

Another strategy for identifying candidate young and compact radio AGN is to measure their radio spectral shapes.  If broadband (or multi-band) radio continuum data are available, young jets may be readily identified by their signature peaked/curved radio spectra, which arise due to absorption.  The radio spectra of most peaked-spectrum (PS) radio sources may be modeled by synchrotron self-absorption (SSA), in which the absorption is caused by the high number density of non-thermal, synchrotron-emitting electrons.  Free-free absorption (FFA), where the absorption is due to a high density of thermal electrons \citep{bicknell+97}, may also produce spectral curvature (e.g.\ in the case of a jet interacting with a dense ionized medium).  By systematically combining measurements from radio surveys spanning a wide range of frequencies, samples of candidate young jets may be constructed (though differences in survey spatial resolutions and observing epochs may complicate the interpretation of such candidates; see \citealt{patil+22}).

The 3rd approach to identifying young jets is the radio time domain.  Compact radio sources have long been known to exhibit variability over a wide range of timescales from seconds to years.  
Among compact extragalactic sources, radio continuum variability may be caused by both extrinsic and intrinsic effects.  Extrinsic variability is caused by propagation effects such as interstellar scintillation \citep{rickett+86, wagner+95}, 
interplanetary scintillation \citep{morgan+18}, and scattering (\citealt{fiedler+87}). Intrinsic radio jet variability may arise from mechanisms such as internal shocks along the jet \citep{marscher+85}, variable Doppler boosting (due to jet re-orientation; \citealt{hovatta+19}), or source evolution (e.g. \citealt{nyland+20}).  Constraints from additional measurements, such as the variability characteristics (e.g.\ the amplitude and timescale) and radio spectral shape, may be used to help distinguish between different effects. 
However, unambiguously determining the underlying cause of radio variability and attributing implied properties such as source compactness to youth (as opposed to effects like relativistic beaming) poses a formidable challenge.  

The best approach for robustly identifying young radio sources is to combine constraints from multiple selection techniques.  
Modern wide-field radio surveys and instruments are being used to identify large samples of PS sources on the basis of their spectral shapes and radio variability.  
Surveys such as the LOFAR Two-metre Sky Survey (LOTSS; \citealt{shimwell+22}) and the GaLactic and Extragalactic All-sky Murchison Widefield Array survey (GLEAM; \citealt{hurley-walker+17}) have had a large impact on the field over the past several years \citep{callingham+17, ross+21, slob+22, ballieux+24}.  

\subsection{Compact Symmetric Objects}
A subset of PS sources are classified morphologically as compact symmetric objects (CSOs; \citealt{peck+00, tremblay+16}).  CSOs are characterized by their symmetric, double-lobed morphologies and their compact ($<$1~kpc) sizes.  They are believed to represent a unique phase in the evolution of a radio AGN in which the jet may begin to ``break out'' of the ISM of the host galaxy \citep{sutherland+07}, and hence are important for understanding the physics of jet-ISM feedback.   
The jet triggering mechanism of CSOs (variable accretion, tidal disruption events, influence of mergers, etc.), as well as the conditions under which they evolve into large-scale radio sources, remain under debate \citep{kiehlmann+24b, readhead+24, sullivan+24}.  

Despite their importance for our understanding of the evolution of radio AGN, the number of confirmed CSO sources is relatively small due to the challenges described above as well as contamination by beamed sources (blazars).  

In this paper, we present new observations taken with the Very Long Baseline Array (VLBA) of a previously unknown CSO candidate that was selected by combining commensal low-frequency radio continuum observations taken with the Karl G. Jansky Very Large Array (VLA) with archival radio data and surveys.  
The data are described in Section~\ref{sec:data}.  We present the results of our observations in Section~\ref{sec:results}.  In Section~\ref{sec:analysis}, we analyze our results. 
We discuss the physical origin of the resolved parsec-scale radio source revealed by the VLBA and the possible implications for jet formation and evolution in Section~\ref{sec:discussion}.   
We summarize the paper and discuss future work in Section~\ref{sec:summary}. 
We adopt a standard $\Lambda$CDM cosmology with $H_{0}$ = 67.7 km~s$^{-1}$ Mpc$^{-1}$, $\Omega_{\Lambda}$ = 0.691 and $\Omega_{\textrm{\small{M}}}$ = 0.307 \citep{planck+15} throughout this paper.  Errors shown represent a 1$\sigma$ uncertainty unless otherwise stated.

\begin{figure*}[t!]
\centering
\includegraphics[clip=true, trim=0cm 0cm 0cm 0cm, width=0.8\textwidth]{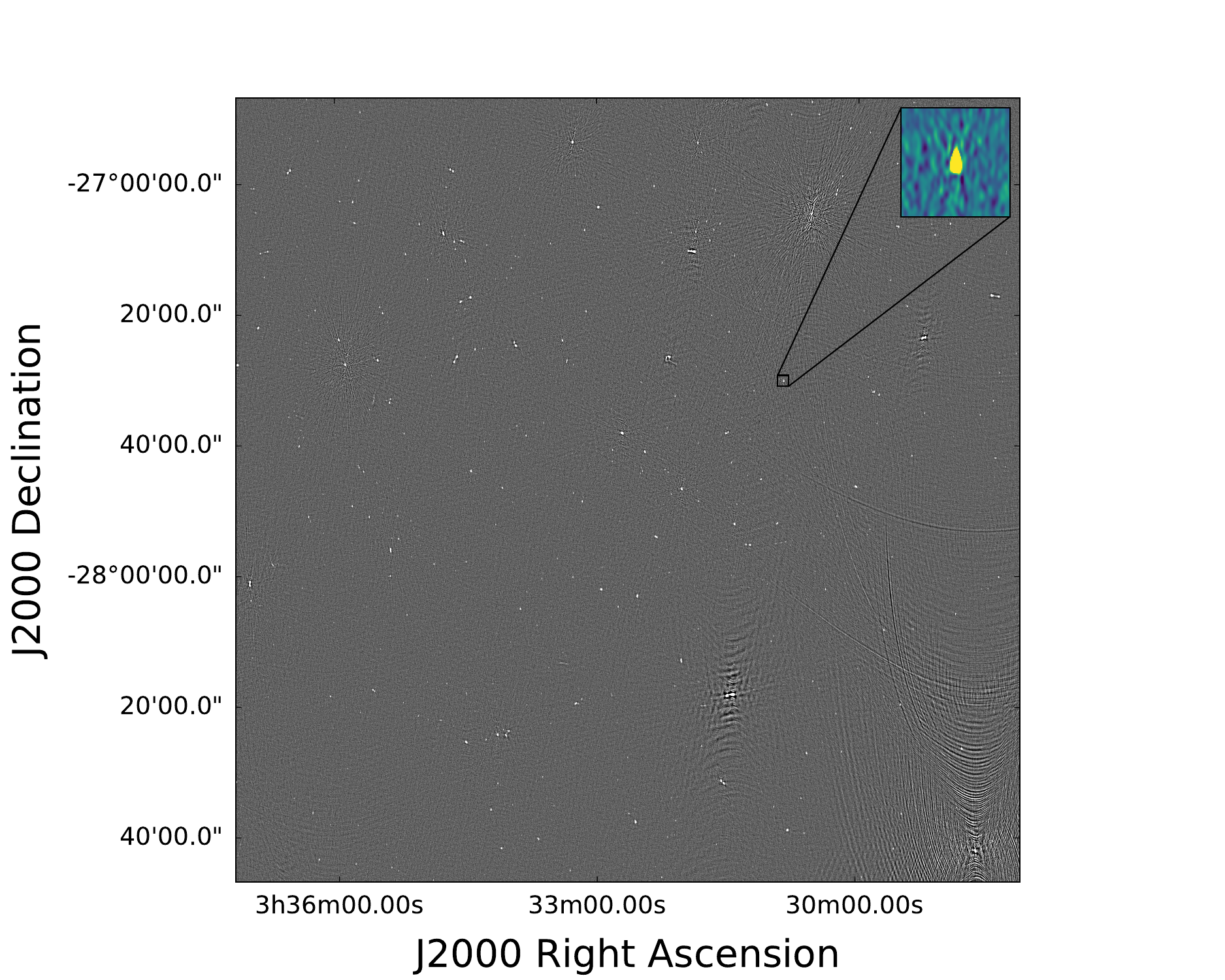}
\caption{
Deep 340~MHz VLITE image centered on the HUDF (Polisensky et al., in prep.). 
The 1$\sigma$ rms noise level is 158~$\mu$Jy~beam$^{-1}$.  The image has dimensions of 2$^{\circ} \times 2^{\circ}$ with  synthesized beam dimensions of $\theta_{\rm maj} \times \theta_{\rm min} = 8.8^{\prime \prime} \times 3.8^{\prime \prime}$. A zoomed-in view of VLITE\_A J033051.4-273014 (J0330-2730) is shown in the 100$^{\prime \prime} \times 100^{\prime \prime}$ inset in the upper right.  
\\
}
\label{fig:HUDF_with_zoombox}
\end{figure*}

\begin{figure}[b!]
\centering
\includegraphics[clip=true, trim=6cm 0cm 7cm 0cm, width=0.425\textwidth]{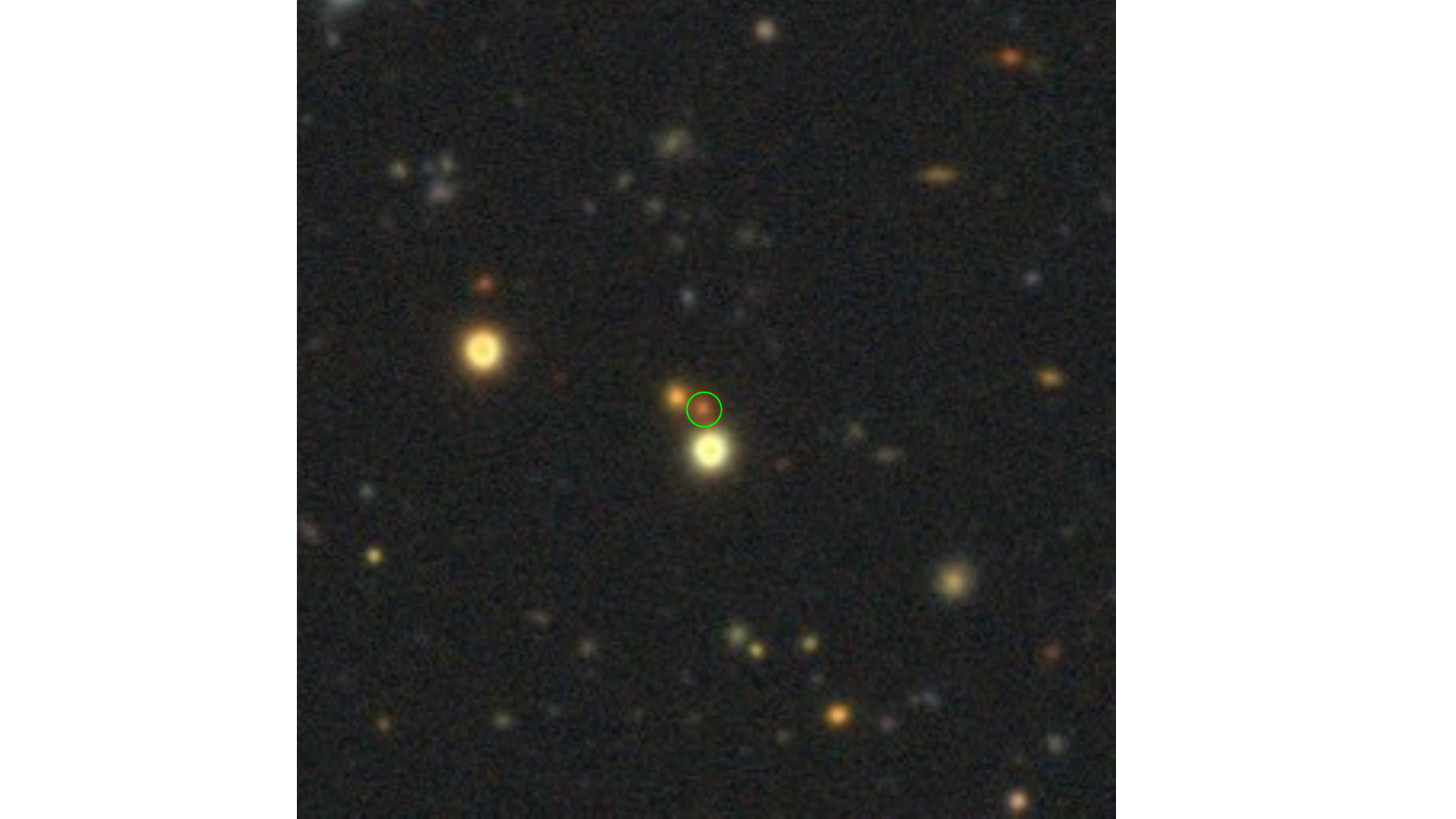}
\caption{Optical griz image cutout from DR10 of the Dark Energy Camera Legacy Survey (DECaLS; \citealt{dey+19}).  The image has dimensions of 65$^{\prime \prime}$ $\times$ 65$^{\prime \prime}$.  The green circle has a diameter of 1$^{\prime \prime}$ and is centered on the VLBA position of J0330-2730. 
\\
}
\label{fig:DECaLS}
\end{figure}

\section{Data} \label{sec:data}

\subsection{VLITE}
The VLA Low-band Ionosphere and Transient Experiment\footnote{\href{http://vlite.nrao.edu}{http://vlite.nrao.edu}} (VLITE) is a commensal system that provides simultaneous data at 340~MHz during regular VLA observing programs \citep{polisensky+16, clarke+16}.  The driving science goals underpinning the development of VLITE include the identification of astrophysical transients and the characterization of the Earth's ionosphere.  VLITE also naturally provides instantaneous information on diffuse/extended structures and point source radio spectral indices.  For extragalactic sources, this information is important for constraining the physical properties of galaxy clusters, AGN, and star-forming galaxies.

VLITE operates on a subset (up to 18) of the 27 VLA antennas with a bandwidth of 38~MHz. The VLITE system includes a dedicated DiFX software correlator, automated data processing pipelines, and a Structured Query Language (SQL) database of cataloged source properties.  The VLITE data are processed using a dedicated calibration and imaging pipeline \citep{polisensky+16} that is based on a combination of the Obit \cite{cotton+08} and AIPS \cite{greisen+03} software packages. 
The flux density scale for VLITE is accurate to $\sim$15\%\footnote{\href{https://cirada.ca/vcsscatalogue}{https://cirada.ca/vcsscatalogue}}. 
Since it began operations in 2014, VLITE has accumulated more than 6200 hours of data per year, making it a vast resource for enhancing the legacy value and scientific impact of regular VLA observations.  


This study is based on VLITE data that were recorded during VLA projects 18A-199 and 19A-242.  The goal of those projects was to produce an ultra-deep (sub-$\mu$Jy-level) 3~GHz image centered on the {\it Hubble} Ultra Deep Field (HUDF; \citealt{beckwith+06}) to study high-redshift AGN and star formation \citep{rujopakarn+16, rujopakarn+18, alberts+20, lyu+22}.  The HUDF was observed at S band (2-4~GHz) in the VLA A configuration from 2018 to 2019 over a series of 45 individual observations with a total integration time of 169 hours.  The commensal VLITE data recorded during these VLA projects provide a unique opportunity to perform time-series analyses and deep imaging at 340~MHz. 
Here, we focus on a small subset of these VLITE observations to explore the commensal radio source population and identify individual sources of interest for further study. 

Out of 45 observations, 6 VLITE datasets were selected for deeper imaging, consisting of 3 observations from 2018 and 3 from 2019.  The 2018 observations were chosen for their high-quality VLITE pipeline images and optimal $uv$-coverage, while the 2019 data were selected to minimize ionospheric distortion.  The data were re-calibrated using 3C48 as the primary calibrator, and the datasets were combined and imaged jointly using WSClean \citep{offringa+14} with joint-channel deconvolution. We also applied phase-only self calibration separately to the 2018 and 2019 datasets. 
The 1$\sigma$ RMS noise levels for the 2018 and 2019 images were 225 and 205 $\mu$Jy~beam$^{-1}$, respectively. The 2018 and 2019 images were then convolved to a common beam of 8.8$^{\prime \prime} \times$ 3.8$^{\prime \prime}$ with a beam position angle of 0$^{\circ}$, and combined in the image plane to produce the final deep image, shown in
Figure~\ref{fig:HUDF_with_zoombox}. 

The final VLITE image consists of 20.66 hours of data and has a 1$\sigma$ RMS noise of $\sigma_{\rm rms} = 158$~$\mu$Jy~beam$^{-1}$.  
Deep imaging and time series analyses of the VLITE data taken during the VLA HUDF observations are in progress, and further details, including a source catalog, will be presented in a forthcoming paper (Polisensky et al., in preparation).  

We combined our deep VLITE image data with measurements taken over a wide range of frequencies using publicly available archival and survey datasets to 
search for PS sources.  A dedicated paper on the radio color selection of PS sources based on VLITE will be presented in a forthcoming study.  
Here, we focus on the properties of a previously unknown Megahertz Peaked Spectrum (MPS) source identified through our deep VLITE imaging, VLITE\_A J033051.4-273014 (hereafter J0330-2730).  This source is located 0.48$^{\circ}$ from the center of the HUDF, well outside the 9$^{\prime}$ primary beam of the original S-band VLA observations but well within the much larger field-of-view of VLITE.  J0330-2730 is shown in the inset box in Figure~\ref{fig:HUDF_with_zoombox}.  The mildly triangular shape of the source is caused by small residual ionospheric errors. J0330-2730 has a counterpart in the optical (Figure~\ref{fig:DECaLS}) and infrared with a redshift based on forced photometry of $z=0.8975$\footnote{At this redshift and given our adopted cosmological parameters from \citet{planck+15}, the luminosity distance to the source is 5945 Mpc.  The angular scale conversion factor is $1^{\prime \prime} = 8$~kpc.} \citep{nyland+23}.  

\begin{deluxetable*}{cccccccc}
\tablecaption{Summary of VLBA Observations for Project BN058 \label{tab:obs}}
\footnotesize
\tablehead{
\colhead{Band} & \colhead{$\lambda$} &	\colhead{Date} & \colhead{Recorded BW}  & \colhead{SPWs} & \colhead{Channels} & \colhead{Backend} & \colhead{Antennas}  \\
\colhead{(GHz)} &  \colhead{(cm)} & \colhead{} & \colhead{(MHz)}  & \colhead{}   & \colhead{} & \colhead{} & \colhead{} 
} 
\colnumbers
\startdata
L & 20  & 2021 Dec 31 & 512  & 8 & 64  & RDBE-PFB & PT, FD, LA, OV, KP, HN, BR, NL, MK, SC \\
S & 13  & 2022 Jan 2  & 512  & 8 & 64  & RDBE-PFB & FD, LA, OV, KP, HN, BR, NL, MK, SC \\
C & 6   & 2022 Jan 11 & 1024 & 4 & 256 & RDBE-DDC & PT, FD, LA, OV, KP, HN, BR, NL, MK, SC \\
\enddata
\tablecomments{Column 1: Receiver band name.  Column 2: Central receiver wavelength.  Column 3: Observing date.  Column 4: Total recorded bandwidth across both polarizations.  Column 5: Number of spectral windows per polarization.  Column 6: Number of channels per spectral window.  Column 7: Digital backend used (RDBE-PFB = polyphase filterbank personality of the Roach Digital Backend; RDBE-DDC = digital downconverter personality).  Column 8: Antennas operational during the observation.  All observations were conducted in dual polarization mode with 2-bit sampling.  PT = Pie Town, FD = Fort Davis, LA = Los Alamos, OV = Owens Valley, KP = Kitt Peak, HN = Hancock, BR = Brewster, NL = North Liberty, MK = Mauna Kea, SC = Saint Croix.}
\end{deluxetable*}

\subsection{VLBA} 
J0330-2730 was observed with the VLBA in the L (20cm), S (13cm), and C (6cm) bands (project ID: BN058). 
The observations were made on 31 December 2021, 02 January 2022, and 11 January 2022, respectively.  Observations at each band were conducted for a total of 4 hours, with 2 hours of on-source time.  The correlator was tuned with an integration time of 1 second.  Phase referencing was performed with the nearby calibrator J0331-2524 with a 2$^{\circ}$ switching angle.  J0324-2918 was used for fringe finding.  A summary of the observations with additional details is provided in Table~\ref{tab:obs}.

The C-band data in this project suffered from substantial gain errors. The main issue stems from incorrect gain values in the FITS-idi files in the NRAO archive.  The archival gain values do not correctly account for gain variations across the full bandwidth of the C-band receivers.  
On the PT and NL stations (see Table~\ref{tab:obs}), a focus rotation problem affecting C-band data taken from 2020-02-11 to 2023-04-05 led to additional gain errors on these antennas.  We followed the official NRAO guidance\footnote{\url{https://science.nrao.edu/facilities/vlba/data-processing/vlba-7ghz-flux-density-scale}} for correcting the C-band gains, which involved downloading NRAO's revised ANTAB table with the correct gains prior to processing the data.  To address the focus errors, we performed careful amplitude self-calibration by first creating a model image of the phase referencing calibrator excluding the problematic antennas, then computing self-calibration solutions based on that model, and finally applying those solutions to the rest of the antennas.  As reported by NRAO, these procedures should yield an absolute flux density scale accuracy of approximately 10\%. 

Our VLBA observations were also affected by severe radio frequency interference (RFI), particularly at S band.  In order to mitigate the RFI, we ran AOFlagger \citep{offringa+10} on each dataset.  All further VLBA data processing, including calibration, additional manual flagging, and imaging, was performed in version 6.5.3 of the Common Astronomy Software Applications (CASA) package \citep{casa+22}.  Since VLBI data processing in CASA is still a relatively new capability, we reviewed recent publications validating the use of CASA for VLBI continuum calibration and imaging \citep{hunt+21, vanbemmel+22}, consulted Memo \#38 in the VLBA memo series\footnote{\protect{\href{https://library.nrao.edu/public/memos/vlba/sci/VLBAS_38.pdf}{https://library.nrao.edu/public/memos/vlba/sci/VLBAS\_38.pdf}}}, and adopted the strategy described in the CASA~6.5.3 VLBA Basic Phase Referencing tutorial available on CASA Guides.\footnote{\protect{\href{https://casaguides.nrao.edu}{https://casaguides.nrao.edu}}}   

\section{Results} \label{sec:results}
We summarize the basic properties of J0330-2730 from our VLITE and VLBA observations in this section.  

\begin{figure*}[ht!]
\centering
\includegraphics[clip=true, trim=0cm 0cm 0cm 0cm, width=\textwidth]{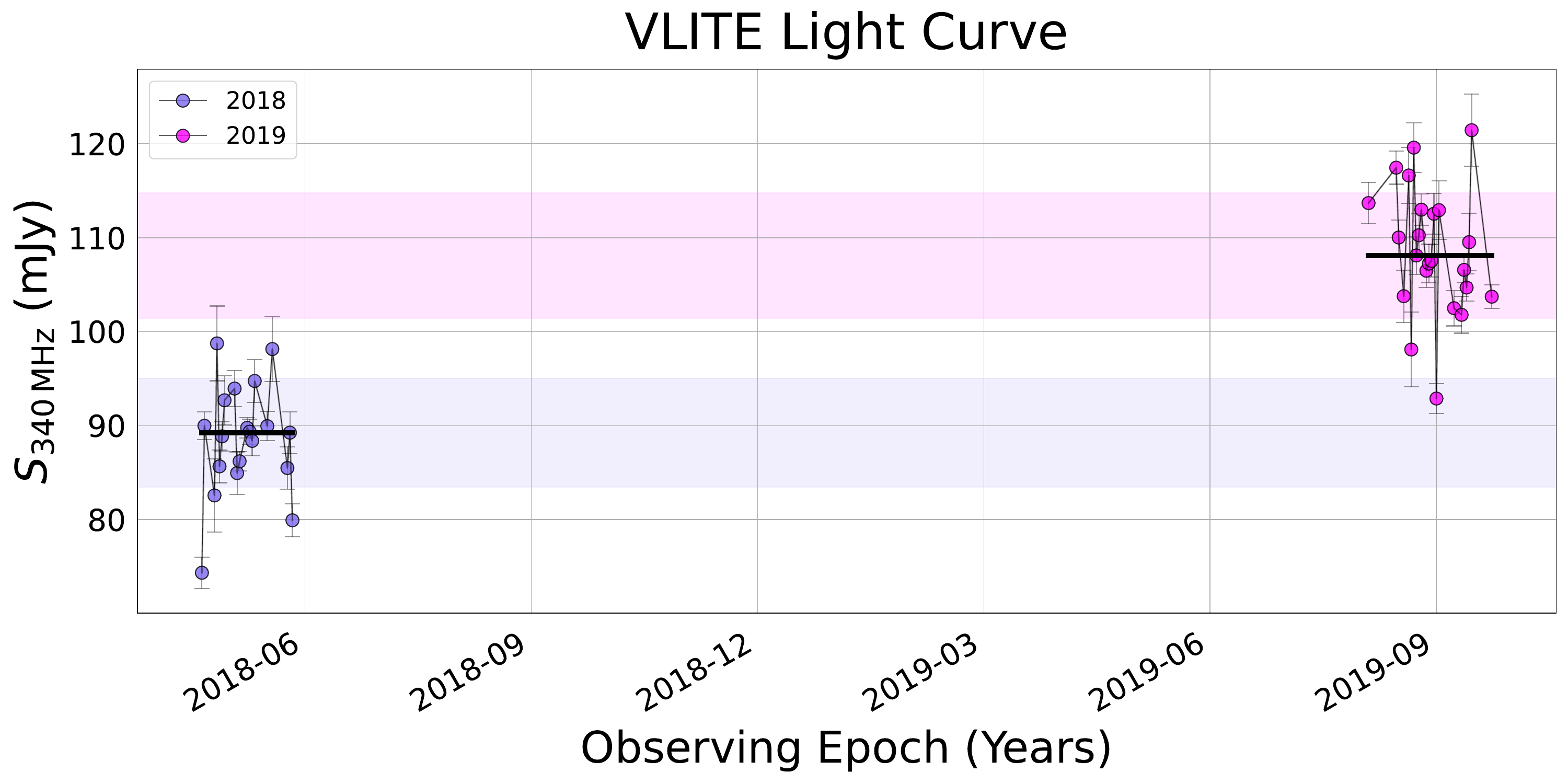}
\caption{The 340~MHz light curve of J0330-2730 from commensal VLITE data obtained during VLA projects 18A-199 and 19A-242, centered on the HUDF. All observations were conducted in the VLA A configuration. Data points from 2018 and 2019 are shown in purple and magenta, respectively, with error bars representing $1 \sigma$ fitting uncertainties (Polisensky et al., in prep.). The median flux densities and standard deviations are 89.26 $\pm$ 5.78 mJy in 2018 and 108.11 $\pm$ 6.68 mJy in 2019, illustrated by the solid black lines and shaded purple and magenta bands.
\\
}
\label{fig:VLITE_light_curve}
\end{figure*}

\subsection{VLITE}
\label{sec:variability}
The total flux of J0330-2730 at 340~MHz as measured in the deep VLITE image shown in Figure~\ref{fig:HUDF_with_zoombox} is 98.40 $\pm$ 2.80~mJy, after primary beam correction \citep{polisensky2024}.  We used this flux to calculate the luminosity following the standard formula (for review, see \citealt{condon+18}) as follows:

\begin{equation}
\label{eq:luminosity}
L_{\rm \nu} = 4\pi D_L^2 (1 + z)^{-\alpha-1} S_{\rm \nu},
\end{equation}
where $L_{\rm \nu}$ is the rest-frame luminosity, $D_L$ is the luminosity distance, $z$ is the redshift, $\alpha$ is the radio spectral index, and $S_{\rm \nu}$ is the observed flux.  For correcting the observed flux to the rest-frame luminosity, we adopt a power-law spectral index of the form $S_{\nu} = a \nu^{\alpha}$, where $S_{\nu}$ is the flux at frequency $\nu$, $a$ represents the amplitude, and $\alpha$ is the spectral index.  
For this calculation, we adopt a flat spectral index ($\alpha=0$) motivated by our radio spectral modeling analysis in Section~\ref{sec:rad_spec_modeling}.  
The radio luminosity of J0330-2730 from our VLITE observation at 340~MHz is therefore 2.19 $\times$ 10$^{26}$ W~Hz$^{-1}$.  

In Figure~\ref{fig:VLITE_light_curve} we show the VLITE light curve of J0330-2730 over approximately one year from 2018 to 2019.  
The images were observed in the VLA A configuration as part of VLA projects 18A-199 and 19A-242.  All measurements were made from images processed by the VLITE imaging pipeline 
and cataloged with the VLITE Database Pipeline \citep{polisensky+19}.  
The fractional variability between the 2018 and 2019 data, calculated as the difference between the maximum and minimum flux divided by the mean, is 20\%. 
This level of variability is larger than the expected flux density scale uncertainty for individual VLITE measurements of 15\%. 
We therefore conclude that J0330-2730 exhibits significant, albeit low-level, flux variability as measured by VLITE.  We further discuss the VLITE variability in  Section~\ref{sec:radio_var} and its physical origin in Section~\ref{sec:vlite_var_origin}. 

\subsection{VLBA}
\begin{figure*}[hbt!]
\centering
\includegraphics[clip=true, trim=0cm 0cm 0cm 0cm, width=0.45\textwidth]{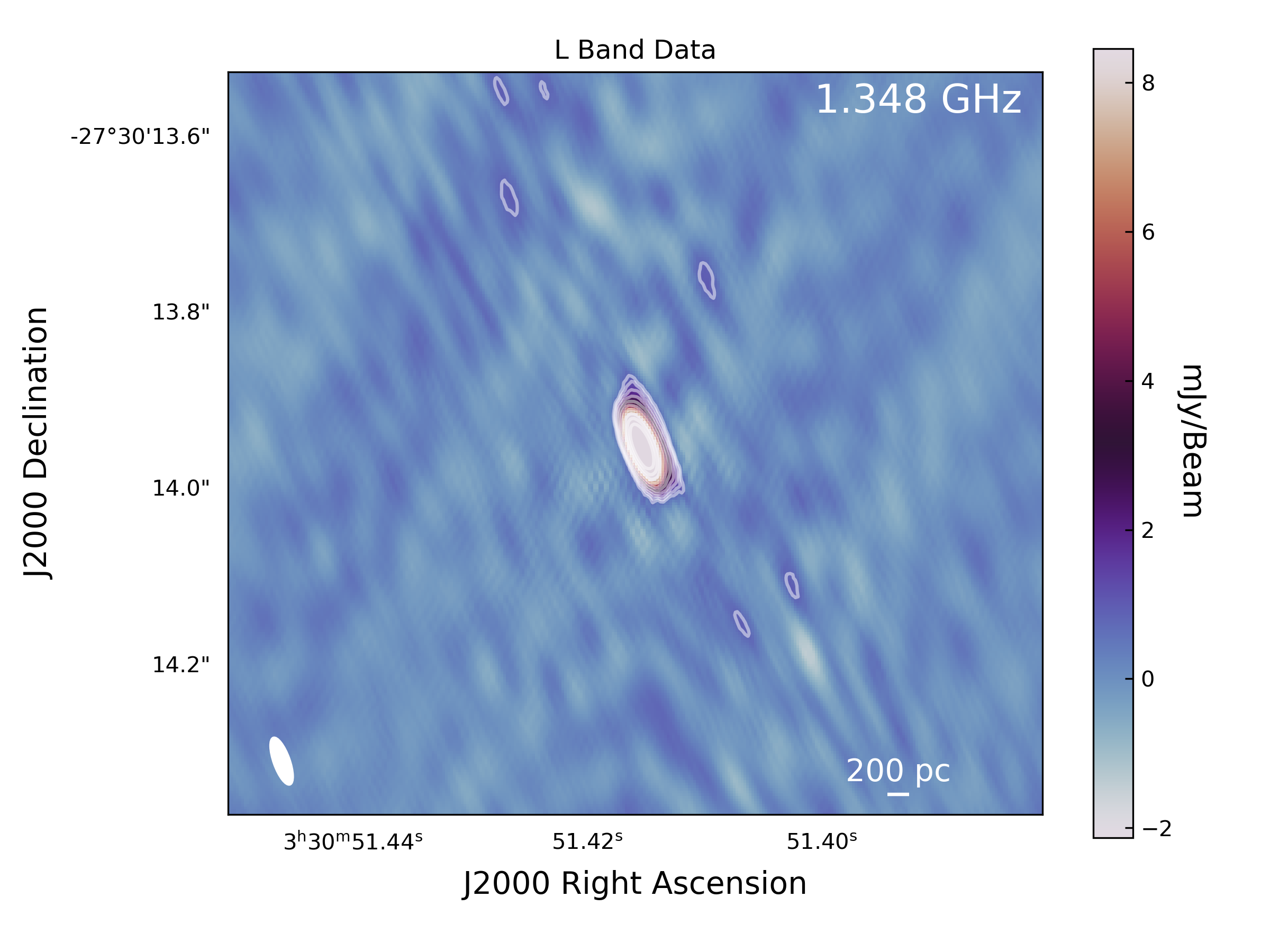}
\includegraphics[clip=true, trim=0cm 0cm 0cm 0cm, width=0.45\textwidth]{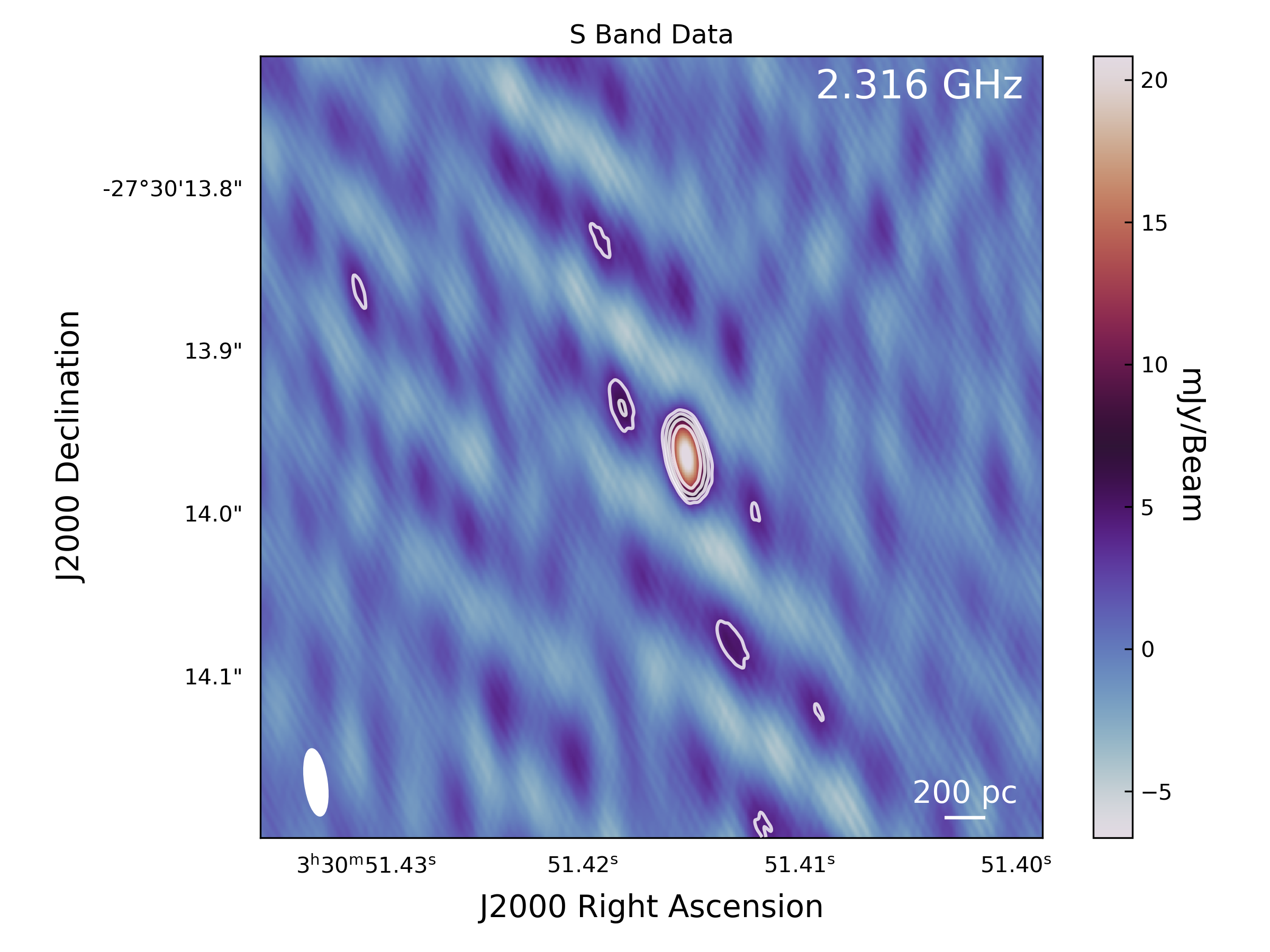}
\includegraphics[clip=true, trim=0cm 0cm 0cm 0cm, width=0.5\textwidth]{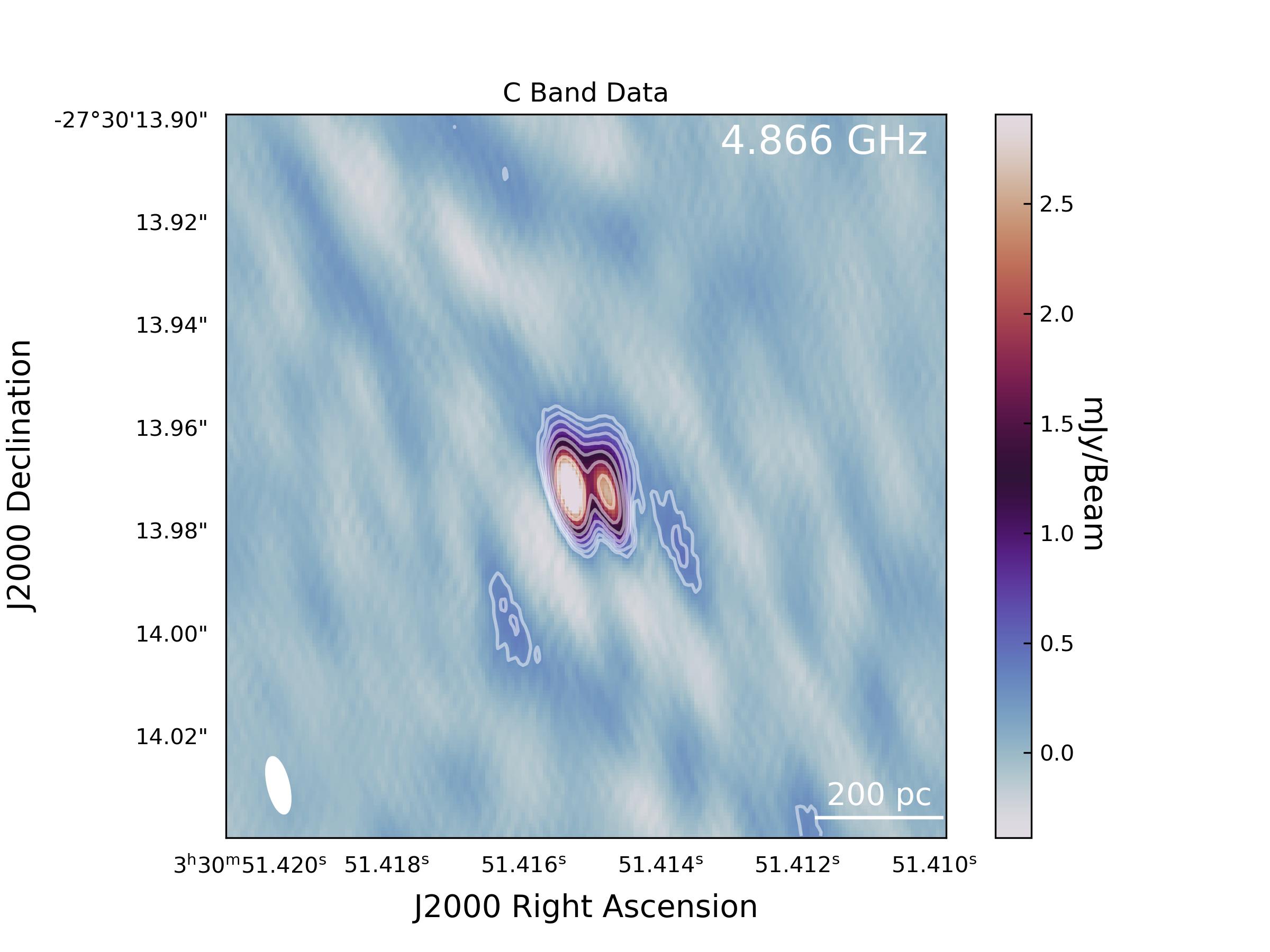}
\caption{VLBA images of J0330-2730 at L, S, and C band.  The central frequency of the image is shown in the upper right corner.  The clean beam is shown as a white ellipse in the lower left corner.  The clean beam dimensions in units of milliarcseconds are 56.30 $\times$ 18.24, 41.00 $\times$ 13.22, 10.64 $\times$ 3.57 at the L, S, and C bands, respectively.  
The contours are shown in intervals of 2$^n$ starting at the 3$\sigma$ level, where 1$\sigma$ represents the RMS noise levels at the L, S, and C bands of 0.28, 1.41, and 0.10 mJy~beam$^{-1}$, respectively.  The image properties are summarized in Table~\ref{tab:VLBA}. 
\\
}
\label{fig:VLBA}
\end{figure*}

VLBA images of J0330-2730 are shown in Figure~\ref{fig:VLBA}. J0330-2730 was detected in all three of the bands included in our VLBA observations.  It has an unresolved morphology at both L and S band, but is resolved into a two-component source at C band.  We discuss the morphology of the source further in Section~\ref{sec:source_classification}.

\begin{deluxetable*}{lccccc}
  \tablecaption{VLBA source measurements \label{tab:VLBA}}
  \tablehead{
  \colhead{} & \colhead{L band} &  \colhead{S band} &  \colhead{C band (left lobe)}  &  \colhead{C band (right lobe)}
  }
  \startdata
Central frequency (MHz) & 1348 & 2316 & \multicolumn{2}{c}{4866}\\
1$\sigma$ RMS noise (mJy beam$^{-1}$) & 0.28 & 1.41 & \multicolumn{2}{c}{0.10} \\
Peak flux density (mJy beam$^{-1}$) & 32.97 $\pm$ 0.28 & 22.18 $\pm$ 1.31 &  6.42 $\pm$ 0.07 & 3.65 $\pm$ 0.07\\
Integrated flux (mJy) & 45.16 $\pm$ 0.84 & 27.57 $\pm$ 2.84 & 6.10 $\pm$ 0.22 & 4.98$\pm$ 0.27\\
log($L_{\nu}$/W Hz$^{-1}$) & 26.20 & 25.99 & 25.33 & 25.25\\
Beam (mas $\times$ mas) &  56.30 $\times$ 18.24 & 41.00 $\times$ 13.22 & \multicolumn{2}{c}{10.64 $\times$ 3.57} \\
Deconvolved major axis (mas) & 15.61 $\pm$ 0.67 &  $<$10.05 & 8.43 ± 0.45 & 11.11 ± 0.81\\
Deconvolved minor axis (mas) & 10.21 $\pm$ 0.12 & \nodata & 2.73 ± 0.07 & 4.28 ± 0.15\\
Linear size (pc)$^{\dagger}$ & 125 & 80 & 93 & 65\\
log($T_{\rm B}$/K) & 8.37 & 8.07 & 6.93 & 6.60 \\
  \enddata
  

 \tablecomments{All errors in this table represent the 1$\sigma$ uncertainty in the two-dimensional elliptical Gaussian fitting performed using the PyBDSF software package.  Upper limits are given at the 3$\sigma$ level. When used elsewhere in the paper for our analysis, all VLBA flux uncertainties include the additional 10\% uncertainty in the absolute flux density scale, added in quadrature to the uncertainties from the fit reported in this table.}
\end{deluxetable*}

Measurements of the source properties, including position, flux, and deconvolved component size, were obtained using PyBDSF \citep{mohan+15}.  PyBDSF is a source extraction software package that may be used to model a source as one or more two-dimensional elliptical Gaussian components.  We used the recommended default parameters in PyBDSF's {\tt process\_image} function.  We report the source properties, including the peak flux density and integrated flux, 
in Table~\ref{tab:VLBA}.  We used Equation~\ref{eq:luminosity} to calculate the VLBA luminosity of each component, but since the observations were made above the turnover frequency at L, S, and C band, we adopted the optically-thin spectral index from our broadband spectral modeling of $\alpha_{\rm thin} = -0.72$ for $\alpha$ in Equation~\ref{eq:luminosity}.  

We also calculated the brightness temperature, $T_{B}$, of each detected VLBA component.  
The brightness temperature of a radio source is defined as: 

\begin{equation}
T_{\mathrm{B}} = \left( \frac{S}{\Omega_{\mathrm{beam}}} \right) \frac{c^2}{2k\nu^2} \left( 1 + z \right),
\end{equation}

\noindent where $S$ is the flux density in units of W~m$^{-2}$~Hz$^{-1}$, $k$ is the Boltzmann constant ($1.38 \times 10^{-23}$~Jy~K$^{-1}$), and $\nu$ is the observing frequency in Hz.  The quantity $\Omega_{\mathrm{beam}}$ is the beam solid angle and is defined as $\Omega_{\mathrm{beam}} = \frac{\pi \theta^2_{\mathrm{FWHM}}}{4 \ln(2)}$, where $\theta_{\mathrm{FWHM}}$ is the angular resolution in units of radians.  The factor of $\left( 1 + z \right)$ accounts for the redshift correction to the source frame brightness temperature.   
The brightness temperature measurements are summarized in Table~\ref{tab:VLBA} and range from $\log(T_{\rm B}/{\rm K}) = 6.60-8.37$.  
These values are consistent with non-thermal synchrotron emission arising from compact components, with no indication of significant relativistic beaming (e.g. \citealt{readhead+94}).

The resolved morphology at C band allows us to constrain the size of the radio source.   
The projected linear size of the radio source is defined as the separation between the peaks of the two components resolved in the C-band VLBA image shown in Figure~\ref{fig:VLBA}.  We measure a projected linear size of $\sim$8~mas, corresponding to a linear size of $\sim$64~pc.   
We compare the linear size of the radio source with predictions from the turnover-size relation (e.g. \citealt{odea+21}, and references therein) in Section~\ref{sec:turnover-size}. 

\section{Analysis} 
\label{sec:analysis}

\subsection{Radio Spectral Modeling}
\label{sec:rad_spec_modeling}
We retrieved data from several public radio surveys spanning 150~MHz to 20~GHz in order to investigate the radio spectrum and morphology of J0330-2730.  Table~\ref{tab:surveys} summarizes the properties of the radio surveys included in this paper.  We note that the angular resolutions of the archival observations span a wide range of values, from 2.5$^{\prime \prime}$ (VLASS) to 100$^{\prime \prime}$ (GLEAM). J0330-2730 has a compact, unresolved morphology in all of the observations listed in Table~\ref{tab:surveys}.  
We performed radio spectral modeling to constrain the evolutionary stage of the radio source.  Prior to analysis, each radio survey cutout image was visually inspected to check for any issues with blending/confusion or poor image quality.  As a result, we rejected the GLEAM single band fluxes as they had low signal-to-noise ratios, keeping only the wideband 200~MHz measurement.  

Radio spectral modeling was performed using the Python code developed in \citet{patil+22} and available to the community on Github\footnote{\faGithub\,\href{https://github.com/paloween/Radio_Spectral_Fitting}{Radio\_Spectral\_Fitting}}. 
Only data from Table~\ref{tab:surveys} were included in the modeling.  The results of our radio spectral modeling analysis are illustrated in Figure~\ref{fig:SED}.  The source has a peaked spectral shape that is well described by either a synchrotron-self absorption (SSA) or free-free absorption (FFA) model with a spectral turnover frequency of 307~MHz (582~MHz in the rest frame of the source at its photometric redshift of $z$=0.8975).  The peak flux density at the turnover frequency of the SSA model fit is 93.4 mJy beam$^{-1}$.  The optically-thin spectral index above the turnover frequency was measured using all data from Table~\ref{tab:surveys} with frequencies at or above 887.5~MHz and has a value of $\alpha_{\rm thin} = -0.72 \pm 0.23$.  The optically-thick spectral index based on TGSS and GLEAM is $\alpha_{\rm thick} = 2.29 \pm 0.75$. 

A potential caveat of our spectral modeling analysis is that the radio spectrum shown in Figure~\ref{fig:SED} is based on fluxes drawn from non-simultaneous multi-band radio data and could therefore be impacted by variability.  Future quasi-simultaneous, multi-band observations will be needed to fully overcome this limitation \citep{nyland+20}.  However, we emphasize that the excellent agreement with the SSA and FFA models suggests that strong variability capable of distorting the spectral shape is unlikely \citep{ross+21}.  We discuss single-band radio variability in more detail in Section~\ref{sec:radio_var}.

We also compare the VLBA fluxes to the broadband radio spectrum of J0330-2730 in Figure~\ref{fig:SED}.  In the case of the C-band measurement, the combined flux of the two components is shown.  The VLBA measurements appear to be consistent with flux measurements at much lower angular resolution within the standard flux uncertainty of 10\% for the VLBA{\footnote{We note that the S-band VLBA flux appears to be slightly low.  This could be due to variability or residual gain errors, possibly associated with the severe RFI that is present at S band.}.  Based on the close agreement between the VLBA fluxes and the archival fluxes at lower angular resolution, we conclude that our VLBA observations have fully captured the radio continuum emission associated with J0330-2730.  In other words, we find no evidence for substantial extended flux (e.g. lobes) missed by the VLBA.

\begin{deluxetable*}{ccccccc}
\centering
\tablecaption{Archival Radio Data \label{tab:surveys}}
\footnotesize
\tablehead{
\colhead{Observation} & \colhead{$\nu$} &	\colhead{$\theta_{res}$} & \colhead{$\sigma_{rms}$}  & \colhead{$S_{\rm peak}$} & \colhead{$S_{\rm total}$} & \colhead{References} \\
\colhead{} &  \colhead{(MHz)} & \colhead{($^{\prime\prime}$)} & \colhead{(mJy/beam)}  &  \colhead{(mJy/beam)} & \colhead{(mJy)} &  \colhead{} 
} 
\colnumbers
\startdata
AT20G       &  20000 &   30  & 0.33 & \nodata  & 6.99 $\pm$ 0.48  & \citealt{franzen+14}\\
ATCA       &  18000 &   10  &  \nodata & \nodata & 7.30 $\pm$ 0.40  & \citealt{franzen+14}\\
ATCA       &  9000  &  25   & \nodata  & \nodata &  12.39 $\pm$ 0.64 & \citealt{franzen+14}\\
ATCA       &  5500  &   40  &  \nodata & \nodata &  20.61 $\pm$ 1.04 & \citealt{franzen+14}\\
VLASS      & 3000   & 2.5 & 0.12 & 24.33 $\pm$ 0.14 & 30.46 $\pm$ 0.22 & \citealt{lacy+20, gordon+21}\\
ATLAS       &  2300  &  57   & 0.08 & \nodata &  35.8 $\pm$ 1.79 & \citealt{zinn+12, franzen+14}\\
ATLAS      & 1395   &  16 & 0.03 & 46.58 $\pm$ 2.38 & 46.58 $\pm$ 2.38 & \citealt{franzen+15} \\
NVSS       & 1400   &  45 & 0.45 & \nodata & 49.10 $\pm$ 1.5  & \citealt{condon+98} \\
RACS-mid   & 1367.5 &  9.7 $\times$ 8.5 & 0.15 & 45.97 $\pm$ 2.76 & 47.38 $\pm$ 2.86 & \citealt{duchesne+24}\\
RACS-low   & 887.5  &  15 & 0.24 &  63.30 $\pm$ 0.24  &  64.92 $\pm$ 5.06 & \citealt{mcconnell+20}\\
VLITE & 340    &  8.8& 0.16 & 83.02 $\pm$ 0.29 & 91.42 $\pm$ 1.03 & This paper\\
TGSS ADR1  & 150    &  25 & 3.5  & 35.45 $\pm$ 5.79 & 46.62 $\pm$ 9.34 & \citealt{intema+17}\\
GLEAM      & 200 & 100 & 6-10 & 84.21 $\pm$ 5.56 & 90.03 $\pm$ 7.00 & \citealt{hurley-walker+17} \\
\enddata
\tablecomments{Column 1: Observation  (survey or telescope acronym).  The acronyms are defined as follows: AT20G = Australia Telescope 20 GHz.  ATCA = Australia Telescope Compact Array.  VLASS = Very Large Array Sky Survey.  ATLAS = Australia Telescope Large Area Survey.  NVSS = NRAO VLA Sky Survey.  RACS = Rapid ASKAP Continuum Survey.  VLITE = VLA Low-band Ionosphere and Transient Experiment.  TGSS ADR1 = TIFR GMRT Sky Survey Alternative Data Release 1.  GLEAM = GaLactic and Extragalactic All-sky MWA survey.  Column 2: Central observing frequency.  Column 3: Angular resolution.  Column 4: Sensitivity, shown as the 1$\sigma$ rms noise.  The value listed for VLASS is for a single epoch.   
We note that no rms noise value is provided for the ATCA observations at 5.5, 9.0 and 18 GHz.  These data were taken as targeted follow-up observations of sources.  
Column 5: Peak flux density and uncertainty.  We note that we have not adjusted the fluxes for differences in underlying flux density scale.  For VLASS, all flux measurements are taken from Epoch 1.  Column 6: Total integrated flux and uncertainty.  The same caveats as for Column 5 apply.    
Column 7: References.  
}
\end{deluxetable*}

\begin{figure*}[t!]
\centering
\includegraphics[clip=true, trim=0.4cm 6.5cm 0.3cm 0cm, width=1\textwidth]{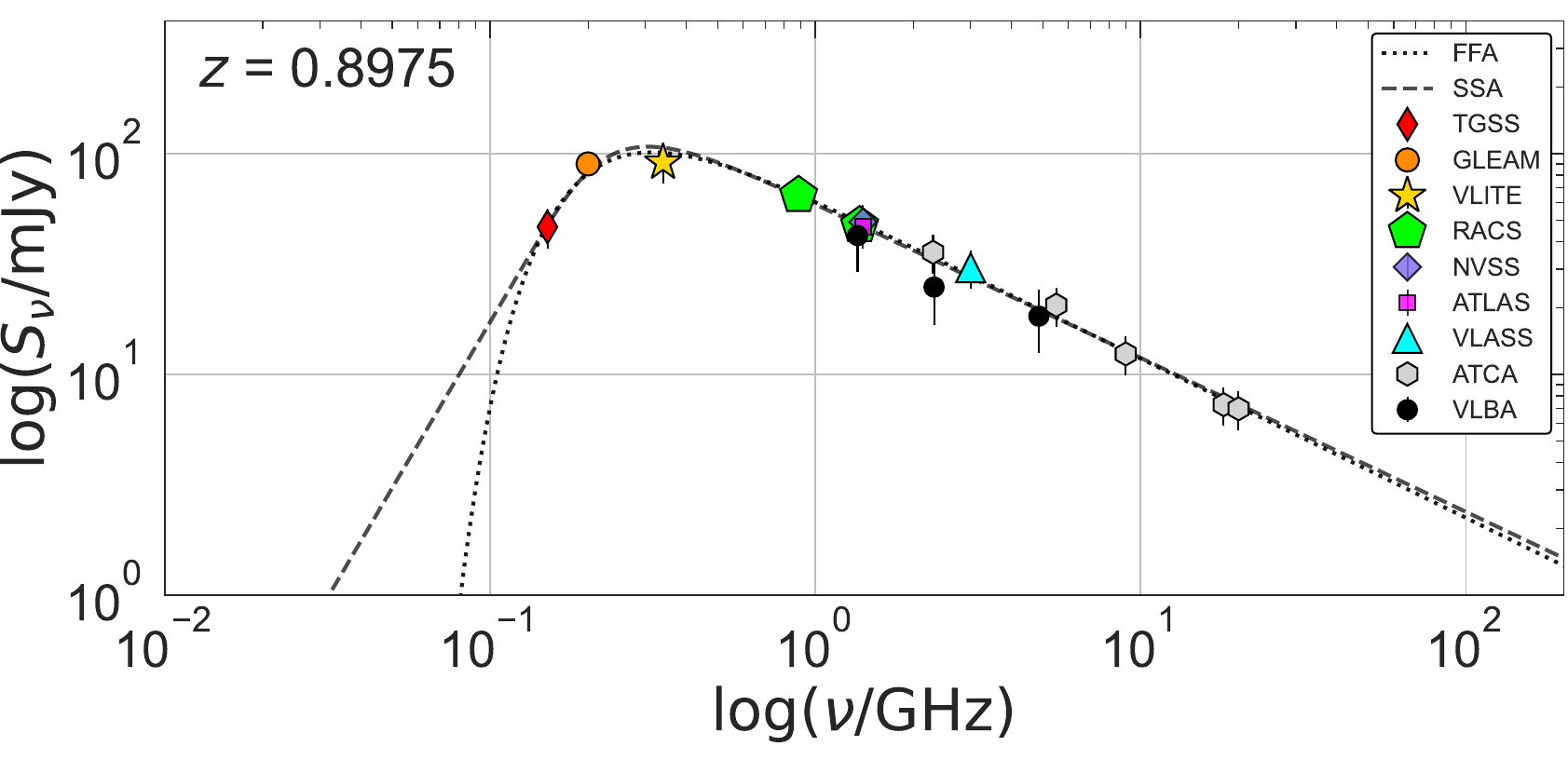}
\caption{Radio spectral modeling 
of J0330-2730.  The different symbols and colors are defined in the legend on the right and correspond to flux measurements from the archival radio surveys listed in Table~\ref{tab:surveys} and our new VLBA observations.  The VLBA fluxes are shown as black points with error bars corresponding to the standard 10\% uncertainty in the VLBA flux density scale.  The source has a peaked spectral shape that is well described by either a synchrotron-self absorption (SSA; dashed line) or free-free absorption (FFA; dotted line) model with a spectral turnover frequency of 307~MHz (582~MHz in the rest frame of the source at its photometric redshift of $z$=0.8975).  The spectral modeling is based on the tools presented in \citet{patil+22}.  The VLBA data points are shown for comparison purposes only and were not included in the spectral modeling.  
\\
}
\label{fig:SED}
\end{figure*}
\subsection{Magnetic Field Estimates}

To further investigate whether SSA is consistent with the observed spectral turnover, we compare magnetic field strengths derived from two independent methods. If SSA dominates, the magnetic field estimated from the SSA turnover properties should be consistent with the equipartition magnetic field, which assumes energy balance between relativistic particles and magnetic fields in the emitting region.  
We estimate the magnetic field strength for the entire source\footnote{Ideally, we would compute the magnetic field strength of the individual source components as in \citep{orienti+08}  However, such component measurements require source flux and size measurements near the turnover frequency.} under the assumption that the spectral turnover arises from synchrotron self-absorption (SSA) as follows: 

{\begin{equation}
    B_{\rm SSA} = \left(  \frac{\nu_{\rm max}}{   f (\alpha) S_{\rm max}^{2/5}  \theta_{\rm mas}^{-4/5}  (1+z)^{1/5}  }  \right)^5, 
\end{equation}

\noindent where $\nu_{\rm max}$ is the turnover frequency in GHz, $S_{\rm max}$ is the total flux at the turnover frequency in Jy, $\theta_{\rm mas}$ is the angular size of the source measured as the projected separation between the peaks of the two components, $z$ is the redshift, and $f(\alpha)$ is a function describing the electron energy distribution that depends weakly on the spectral index \citep{kellerman+81}.  
We adopt a standard value of 8 for $f(\alpha)$ \citep{odea+98}.  
Given these measurements and assumptions, we estimate a magnetic field strength for the entire source of $\sim20$ mGauss.

For comparison, we also estimated the magnetic field strength for the case of equipartition between the particle and magnetic field energy densities as follows \citep{miley+80, patil+22}:

\begin{equation}
\label{eq:Bmin}
    B_{\rm min} = 0.0152 \left(  \frac{a}{f_{\rm rl}}  \frac{(1+z)^{4-\alpha}}{\theta_{\rm mas}^3}  \frac{S}{\nu_{\rm GHz}^{\alpha}} \frac{X_{0.5}(\alpha)}{r_{\rm Mpc}} 
 \right)^{2/7}.   
\end{equation}

\noindent In Equation~\ref{eq:Bmin} above, $a$ is the relative contribution of the ions to the total energy, $f_{\rm rl}$ is the filling factor, $\theta_{\rm mas}$ is the linear size of the source, $S$ is the flux density measured at optically-thin frequency $\nu_{\rm GHz}$, 
$r_{\rm Mpc}$ is the co-moving distance at the redshift of the source, and $X(\alpha)$ defines the frequency range over which the radio spectrum is integrated, defined as:

\begin{equation}
    X(\alpha) = \frac{(\nu_{2}^{p+\alpha} - \nu_{1}^{p+\alpha}  )}{(p+\alpha)},
\end{equation}

\noindent where $p$ is the electron energy distribution and $\alpha$ is the optically-thin spectral index.  For the $B_{\rm min}$ estimates we adopt the total VLBA flux measured at 4.866~GHz of 18.38~mJy for $S$.  We set $\theta_{\rm mas}$ to 8~mas, which is the projected linear separation between the lobes measured in our VLBA C-band image.  
The remaining parameters were set to $f_{\rm rl} = 1$, $a=2$, $\alpha = -0.72$, $\nu_{1} = 0.01$~GHz, $\nu_{2} = 100$~GHz, and $p=2\alpha+1=-0.44$.   
We estimate an equipartition magnetic field strength of 
$B_{\rm min} \approx10$ mGauss}. 
 
The $B_{\rm min}$ values are consistent with the magnetic field strengths of CSOs reported in the literature (e.g. \citealt{orienti+08}).  The rough agreement between the $B_{\rm SSA}$ and $B_{\rm min}$ values indicates that the source is consistent with equipartition expectations and a SSA origin for the absorption.  

\subsection{Radio variability}
\label{sec:radio_var}
Variations in the radio fluxes of AGN and quasars are common and provide important insights into the physical conditions of jets and their environments (e.g., \citealt{barvainis+05, mooney+19, bell+19, ross+21, nyland+20}).
In the GHz domain, blazars often exhibit substantial variability in their fluxes and spectral shapes due to relativistic beaming effects, which helps distinguish them from young sources with small inclination angles to our line of sight (e.g., \citealt{orienti+20, hovatta+19}). Long-term monitoring campaigns have shown that blazar variability amplitudes typically exceed 40\% above 1~GHz, with characteristic timescales spanning months to years (e.g. \citealt{richards+11}), although the time domain behavior varies by their sub-classification (e.g. \citealt{liodakis+17}).  

We assess whether the moderate variability observed by VLITE for J0330-2730 ($\sim$20\%~yr$^{-1}$ fractional variability) 
provides a meaningful constraint on its classification as a CSO, as defined by \citet{kiehlmann+24}. We provide a more comprehensive discussion of the possible physical origins of the VLITE variability in Section~\ref{sec:vlite_var_origin} and focus here on CSO variability.

\begin{figure}[h!]
 \centering
 \includegraphics[clip=true, trim=0cm 0cm 0cm 0cm, width=0.45\textwidth]{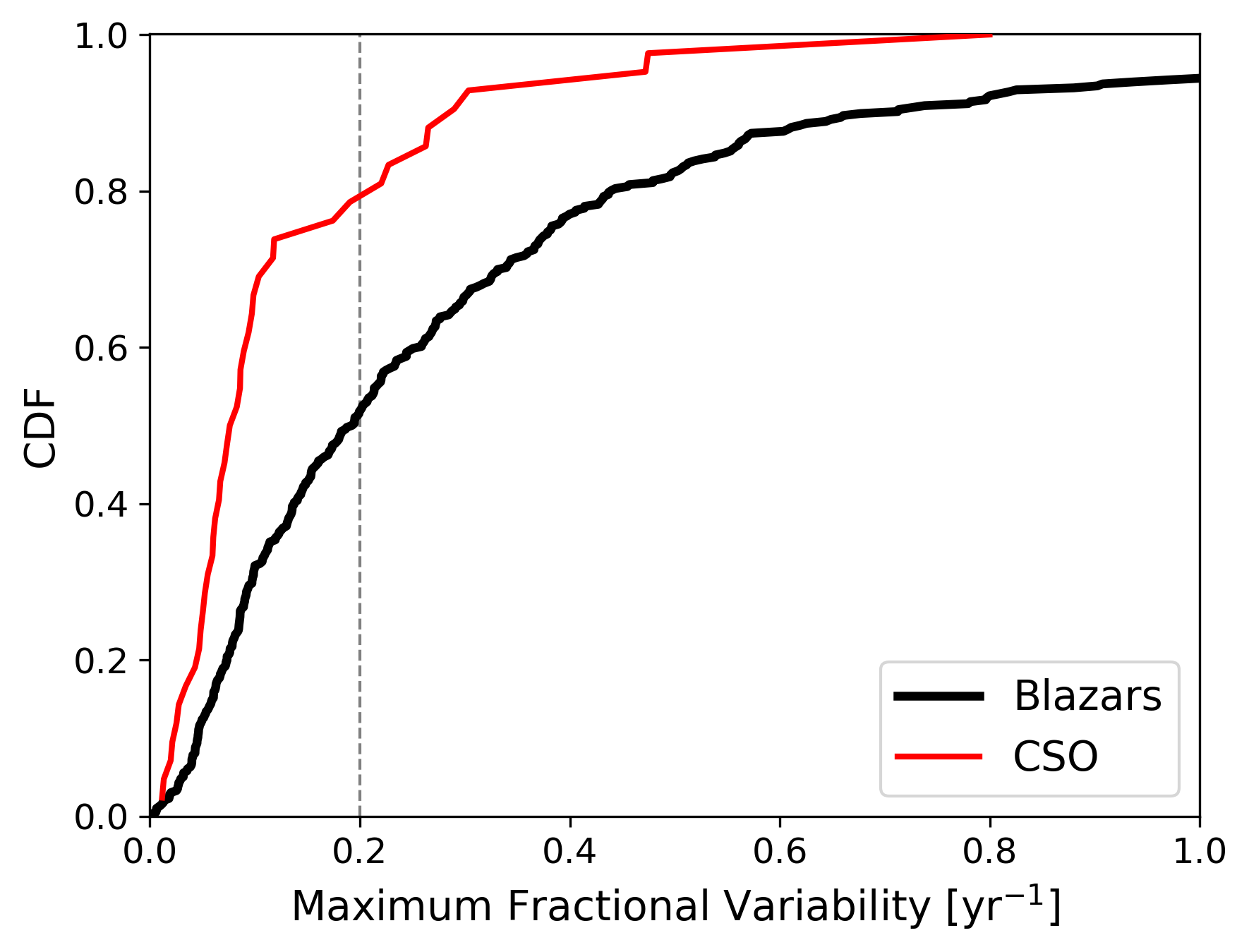}
 \caption{Cumulative distribution function of the VLITE 340~MHz maximum fractional variability per year for a sample of CSOs selected from \citet{kiehlmann+24} and \citet{sheldahl+25} plus blazars from the RomaBzCat \citep{massaro+15}.  Dashed line gives the CSO selection criterion at GHz frequencies ($< 20\%$~yr$^{-1}$) of \citet{kiehlmann+24}. The VLITE light curves have been filtered to remove spurious values as described in Section~\ref{sec:radio_var}.  
 \\
 }
 \label{fig:var_compare}
 \end{figure}

Variability characteristics across the CSO and blazar populations have not yet been thoroughly explored at low radio frequencies. To establish a baseline for VLITE observations, we analyzed the light curves of a combined sample of CSOs from \citet{kiehlmann+24} and \citet{sheldahl+25} alongside a comparison sample of  
blazars from the fifth edition of the RomaBZcat catalog \citep{massaro+15}. Only sources with redshifts reported in these publications were included in the analysis.  To mitigate the impact of outliers and sparsely sampled data, we applied a median filter within a $\pm180$ day window around each data point, adopting the median flux density when at least ten measurements were available. 

To quantify flux density variations, we used the maximum fractional variability metric from \citet{kiehlmann+24}:
\begin{equation}
V = \frac{\Delta S}{S_{\rm min}}  \frac{1+z}{\Delta t}
\end{equation}
where $\Delta S$ is the difference between the maximum and minimum flux density $S_{\rm min}$, $\Delta t$ is the timespan between the flux extrema, and $z$ is the redshift. We selected sources with cataloged redshifts and filtered light curves spanning $>1$ year, yielding a final sample of 396 blazars and 42 CSOs.  

Figure~\ref{fig:var_compare} presents the cumulative distribution of variability for both samples. While the CSO sample is small, CSOs exhibit significantly lower variability than blazars based on this metric. The $< 20\%$~yr$^{-1}$ variability criterion of \citet{kiehlmann+24}, originally defined at GHz frequencies, also appears to be effective at 340 MHz: $\sim$80\% of CSOs fall below this threshold, compared to only $\sim$50\% of blazars. Although variability alone is insufficient to classify an individual source definitively, this trend provides statistical support for applying the \citet{kiehlmann+24} variability criterion in Section~\ref{sec:source_classification}.

\begin{figure*}[t!]
 \centering
 \includegraphics[clip=true, trim=0cm 0cm 0cm 1.25cm, width=0.95\textwidth]{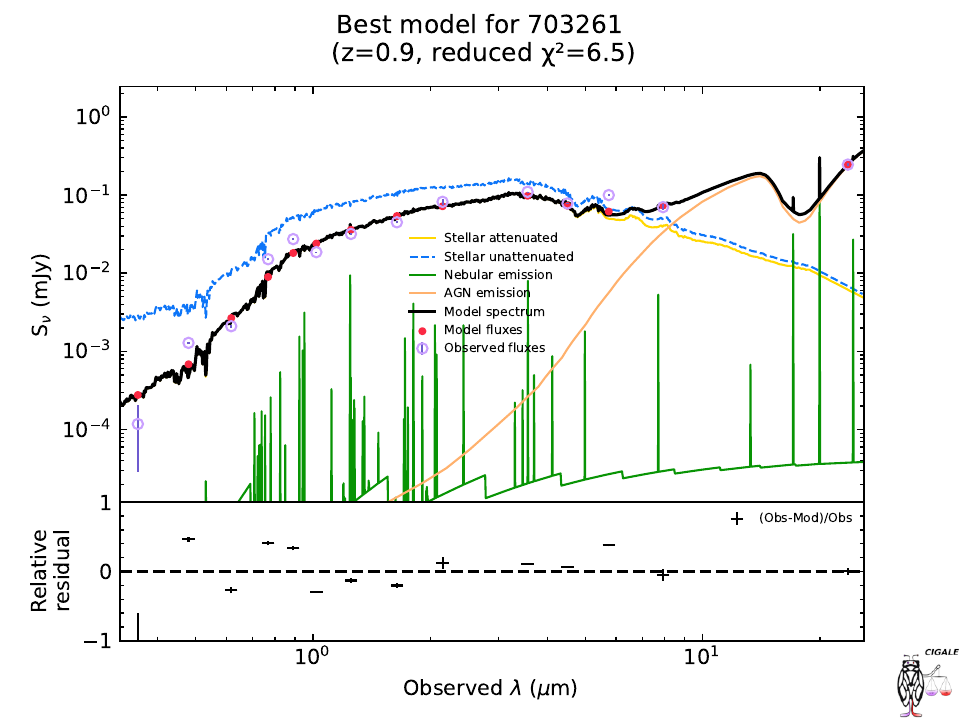}
 \caption{SED modeling of J0330-2730 with CIGALE.  The measured fluxes are the purple open circles with errors. The red circles are the predicted model fluxes at the wavelength of the observed fluxes. The model components are shown by the lines, as indicated in the legend.  
 The fit has a reduced $\chi^{2}=6.5$. 
 This figure shows that the SED of J0330-2730 is consistent with a massive galaxy harboring a low-luminosity AGN, as discussed in Section~\ref{sec:host}.  
 \\
 }
 \label{fig:cigale}
 \end{figure*}
 
\label{sec:host}
In order to investigate the basic properties of the host galaxy of J0330-2730, we compiled optical and infrared data from publicly available archives.  
Specifically, we used the forced photometry from \citet{nyland+23}, which includes the 3.6 and 4.5$\mu$m {\it Spitzer}/DeepDrill survey \citep{lacy+21} bands; ground-based near-infrared data in the $Z$, $Y$, $J$, $H$, and $Ks$ bands from the VISTA Deep Extragalactic Observations (VIDEO; \citealt{jarvis+13}) survey; and optical data in the $g$, $r$, $i$, and $z$ bands from the Hyper Suprime-Cam instrument on the Subaru telescope \citep{Ni+19}. In addition to this photometry, we also included data from Sloan Digital Sky Survey (SDSS; \citealt{york+00}) in the u, r, and i bands as well as archival {\it Spitzer} data in the IRAC3 (5.7~$\mu$m), IRAC4 (7.9$\mu$m), and MIPS (24$\mu$m) bands from the {\it Spitzer} Enhanced Data Products \citep{seip+20}.  

To gain further insights into the properties of the AGN and its host galaxy, we used the optical and infrared fluxes to perform spectral energy distribution (SED) modeling using the Code Investigating GALaxy Emission (CIGALE; \citealt{boquien+19}).  We assumed the BC03 stellar population synthesis model \citep{bc03}, a Salpeter initial mass function, and the SKIRTOR2016 AGN model \citep{skirtor2012,skirtor2016}. 

Figure~\ref{fig:cigale} shows the results of our SED modeling with CIGALE.  Our model results indicate that the galaxy has a mass of $2.6\times 10^{11}$ $M_{\odot}$ and the AGN has a luminosity of $2.0\times 10^{11} L_{\odot}$, consistent with typical Seyfert galaxies \citep{ho+08}. We emphasize that the SED modeling shown here is based on sparse photometric measurements and is primarily intended to serve as a guide for future spectroscopic observations that will enable more robust measurements of the galaxy mass and AGN luminosity.  

\subsection{SMBH Mass Estimate}
The mass of the SMBH is a fundamental parameter that plays a significant role in the formation and evolution of jets (e.g.\ \citealt{ho+02, blandford+19}).  
We used the stellar mass from our SED modeling analysis described in Section~\ref{sec:host} to obtain a rough estimate of the SMBH mass using the 
$M_{\star}-M_{\rm SMBH}$
relation \citep{magorrian+98, ferrarese+00, kormendy+13}.  Our estimate is based on the relationship for low-redshift AGN from \citet{reines+15}, which found a SMBH-to-total stellar mass fraction of 0.025\%.  Assuming a stellar mass of $4\times10^{11}$ M$_{\odot}$, the SMBH mass is expected to be approximately $\sim$10$^8$ M$_{\odot}$.  

\section{Discussion} 
\label{sec:discussion}
We discuss the radio properties of J0330-2730 and their implications.  Our discussion focuses on the origin of the VLITE variability, the classification of the source, and the implications for jet evolution.

\subsection{Origin of the VLITE Variability}
\label{sec:vlite_var_origin}
Our observations reveal significant ($\sim$20\%) fractional variability at 340~MHz between epochs separated by 1.3 years for J0330-2730.  Here we evaluate three potential mechanisms that could explain the observed variability: refractive interstellar scintillation, variable free-free absorption, and relativistic beaming.

\subsubsection{Refractive Interstellar Scintillation}
Below 1~GHz, slow radio variability with an amplitude of 10-30\% on timescales of months to years may arise from refractive interstellar scintillation (RISS; \citealt{rickett+86}).  RISS occurs when radio waves from compact sources are scattered by electron density fluctuations in the interstellar medium \citep{rickett+90}.  Two important parameters that influence the variability amplitude and timescale due to RISS are the observing frequency and Galactic latitude of a source.  
Given the Galactic latitude of our target of $b = -54.7438^{\circ}$, the critical frequency, $\nu_{o}$, below which the modulation due to scintillation is in the strong regime, is $\approx$6~GHz (from Figure~1 of \citealt{walker+98}).  Our VLITE observations are therefore well within the strong scattering regime. 
Following \citet{walker+98}, the expected flux modulation at 340~MHz is m = ($\nu$/$\nu_0$)$^{17/30}$ $\sim$ 20\%. This modulation is expected to occur on a timescale of $t_r \sim 2(\nu_0/\nu)^{11/5}$ hours $\sim$ 46 days.

The observed 20\% difference between our two VLITE measurements is consistent in amplitude with the 20\% modulation predicted for RISS at 340~MHz. However, given the sparse sampling presented in this paper (two epochs separated by 1.3 years), we cannot robustly constrain the variability timescale.  Nonetheless, RISS remains a plausible explanation. 
Future analysis of a more well-sampled VLITE light curve will provide better constraints on the nature of the 340~MHz variability of this source.

\subsubsection{Variable Free-Free Absorption}
\label{sec:ffa}
The spectral turnover at 307~MHz indicates the presence of absorption, potentially due to FFA from a circumnuclear ionized medium \citep{bicknell+97}. Variable FFA presents a plausible explanation given that our observations at 340~MHz are very close to the turnover frequency, where optical depth effects are most prominent. 
The optical depth for free-free absorption follows: 
\begin{equation}
\tau_{\text{ff}} \approx 0.082 \left(\frac{n_e}{\text{cm}^{-3}}\right)^2 \left(\frac{T}{10^4~\text{K}}\right)^{-1.5} \left(\frac{\nu}{\text{GHz}}\right)^{-2.1} \left(\frac{L}{\text{pc}}\right)
\end{equation}
where $n_e$ is electron density, $T$ is temperature, $\nu$ is frequency, and $L$ is path length through the absorbing medium \citep{odea+98}. 
For a spectral turnover at $\nu \approx 307$ MHz (observed frame), where $\tau_{\text{ff}} \approx 1$, even modest changes ($\sim$10\%) in the properties of the ionized medium could produce the observed 20\% flux variations.  This can be understood by considering the relationship between observed flux and the free-free optical depth (e.g. \citealt{rybicki+79}):
\begin{equation}
S_{\nu} \approx S_{\nu,\text{intrinsic}} \cdot e^{-\tau_{\text{ff}}}.
\end{equation}
For $\tau_{\text{ff}} \approx 1$, a fractional change in optical depth of $\Delta\tau_{\text{ff}}/\tau_{\text{ff}} \approx 0.1$ would result in a flux ratio:
\begin{equation}
\frac{S_2}{S_1} \approx \frac{e^{-0.9}}{e^{-1}} \approx 1.22,
\end{equation}
which closely matches our observed 20\% variability.

Timescales of 1-2 years in the observed frame for changes in FFA are physically reasonable for circumnuclear environments in AGN. Similar variability timescales have been found for sources exhibiting variable FFA absorption in the literature (e.g. \citealt{tingay+03, tingay+15}).  

Variable FFA has interesting implications for our understanding of the importance of jet-driven feedback for galaxy evolution.  The free electrons responsible for absorbing the radio emission from the jet in FFA may originate from a foreground screen or arise from jet-driven heating or shock formation \citep{vermeulen+03, zovaro+19}.  Evidence for such jet-driven interactions may include multi-phase outflows (e.g. \citealt{alatalo+11, nyland+13, mukherjee+16, murthy+24}). Young radio sources with evidence for variable FFA are therefore good candidates for future follow-up studies aimed at characterizing the gas content and conditions in the vicinity of the jets/lobes.  




\subsubsection{Relativistic Beaming} 
Given the mild asymmetry in the lobes reported in Section~\ref{sec:results}, we consider whether relativistic beaming could have a significant impact on the VLITE variability of our source. Comparing the integrated fluxes of the two VLBA components reported in Table~\ref{tab:VLBA}, we find a flux ratio of:
\begin{equation}
    R = S_{j}/S_{cj} = 6.10/4.98 \approx 1.22,
\end{equation}
\noindent where $S_{j}$ and $S_{cj}$ are the jet and counterjet fluxes, respectively. Here, we assume that the brighter lobe component is the jet and the fainter one is the counterjet.  

We then used the flux ratio of the lobes to estimate the inclination angle\footnote{The inclination angle, $\theta_i$, is defined as the angle between the jet axis and the line of sight such that $\theta_i = 90^{\circ}$ corresponds to jets aligned perfectly in the plane of the sky.} of the jets assuming that the Doppler boosting is the cause of the asymmetries in the resolved morphology of the source using the following standard equation (e.g. \citealt{wezgowiec+24}, and references therein):

\begin{equation}
    \theta_i = \left[ {\rm acos} \left( \frac{1}{\beta} \frac{(s - 1)}{(s + 1)} \right) \right].   
    \label{eq:inclination}
\end{equation}
In Equation~\ref{eq:inclination}, $\beta$ is the jet velocity in units of the speed of light, $c$, and $s = (S_{j}/S_{cj})^{(1/2 - \alpha)}$.
The jet velocity of our source is unknown, so we adopted a moderately relativistic jet speed with $\beta$ = 0.5 \citep{arshakian+04}.  This speed is on the upper end of typical values for CSOs, which have been found to range from $\beta = 0.1-0.4$ \citep{an+12}.  
We adopt the optically-thin spectral index of $\alpha=-0.72$.  
With these assumptions, we estimate an inclination angle of $\theta_i \approx 75^{\circ}$.  

A small change in either the jet orientation or its velocity may therefore lead to flux variability.  Following \citet{urry+95}, for a continuous jet, and using our convention for $\alpha$ defined by $S \sim \nu^{+\alpha}$), the observed flux ($S$) relates to the intrinsic flux ($S_0$) as:
\begin{equation}
S = S_0 \times \delta^{2-\alpha}, 
\end{equation} 
where $\delta$ is the Doppler factor.  The Doppler factor is defined as:
\begin{equation}
\delta = \frac{1}{\gamma(1-\beta\cos\theta_i)},
\end{equation}
\noindent where $\gamma = \frac{1}{\sqrt{1-\beta^2}}$ is the Lorentz factor.  
For a source with a spectral index of $\alpha = -0.72$, a 20\% flux increase would require:
\begin{equation}
\frac{S_2}{S_1} = 1.2 = \left(\frac{\delta_2}{\delta_1}\right)^2 \rightarrow \left(\frac{\delta_2}{\delta_1}\right) = 1.069 \approx 7\%
\end{equation}
For a mildly relativistic jet with our adopted value of $\beta=0.5$ and an orientation of $\theta_i = 75^{\circ}$, a 7\% change in the Doppler factor could be achieved with a decrease\footnote{We note that increasing the orientation angle would decrease the Doppler factor leading to relativistic deamplification.} in the jet orientation angle by $\approx 5-10^{\circ}$. 

While VLBA monitoring studies of blazars have detected jet position angle changes of several degrees per year \citep{lister+13, punsly+21}, such large changes in orientation angle for lobe-dominated sources like CSOs are difficult to explain.  Significant changes in CSO structures with lobe extents of 10's to 100's of pc require decades to centuries due to light-travel time constraints (e.g. \citealt{an+12, tremblay+16, orienti+20}).  Assuming that the majority of the 340~MHz emission originates from the lobes and not an unidentified core (which is supported by the broadband radio spectrum shown in Figure~\ref{fig:SED}), we conclude that rapid orientation changes on timescales of a year are 
not a plausible origin for the VLITE variability in J0330-2730.

\subsubsection{Distinguishing Between Mechanisms}
RISS, variable FFA, and relativistic beaming can all plausibly produce variability with a similar magnitude and timescale at 340~MHz.  Distinguishing between these effects will require more detailed measurements of the frequency dependence of the emission and the long-term evolution of its spectral shape.   

To investigate the frequency dependence in more detail, we also checked for variability at higher radio frequency across the three epochs of VLASS at 3~GHz.  VLASS provides high-resolution (2.5$^{\prime \prime}$), multi-epoch observations of the radio sky with a cadence of 32 months between epochs.  J0337-2730 was observed as part of VLASS in 2018, 2020, and 2023.  Using the available quick-look VLASS image products \citep{lacy+20}, we measured the peak flux density at each epoch and calculated a fractional variability of $\sim$10\%.  We note that this value is within the expected level of flux uncertainty for the quick-look image products \citep{gordon+21}.  
We therefore find no evidence for significant radio variability at 3~GHz based on VLASS data.  

Stronger variability at lower frequency is roughly consistent with either RISS or variable FFA, but we caution that our VLITE and VLASS variability measurements are not well sampled in either time or frequency, and  systematic multi-frequency observations will be needed to firmly distinguish between these mechanisms.





\subsection{Source Classification}
\label{sec:source_classification}

\begin{figure*}[t!]
\centering
\includegraphics[clip=true, trim=0cm 0cm 0cm 0cm, width=0.85\textwidth]{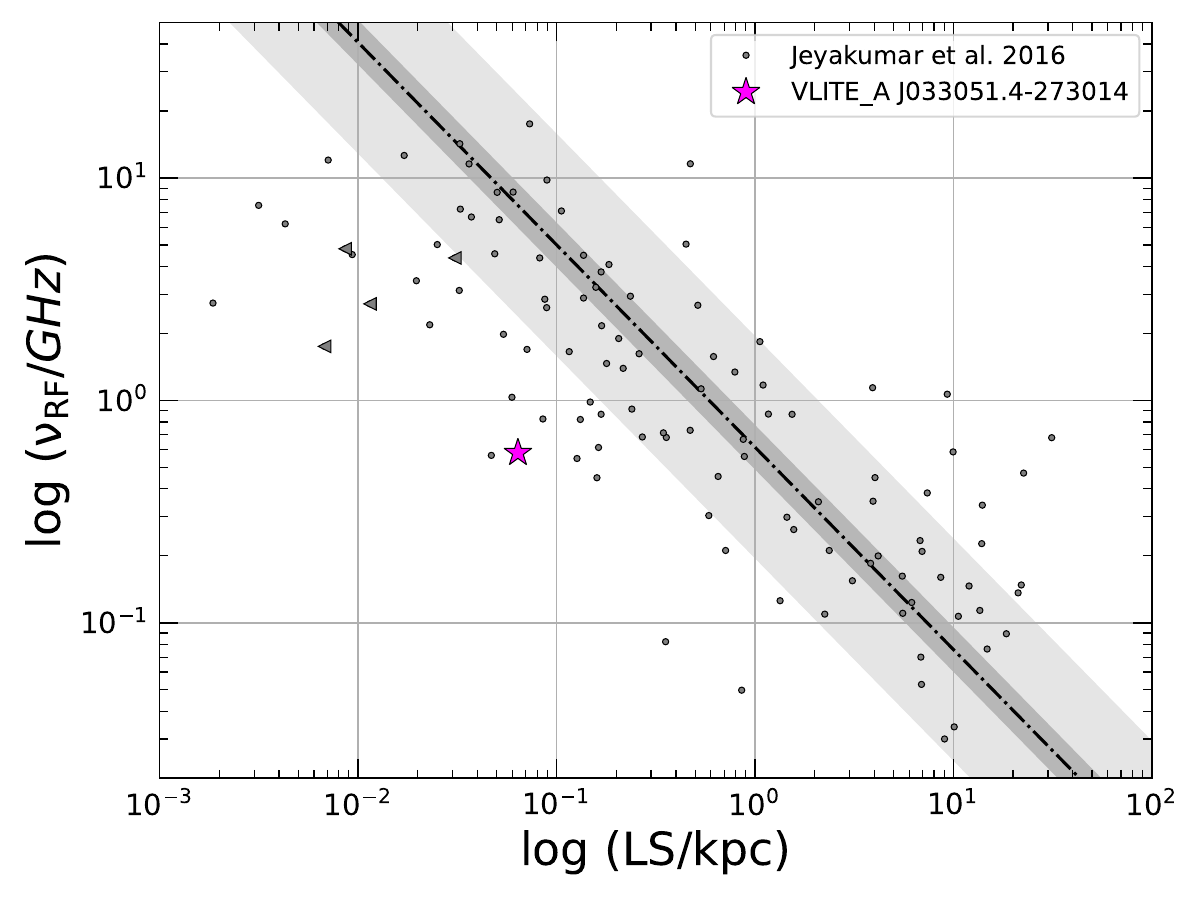}
\caption{Spectral turnover vs.\ linear size for a sample of peaked spectrum sources from the literature \citep{jeyakumar+16}.  The triangles represent sources with upper limits to their sizes.  The fit to the relation from \citet{odea+97} is shown by the black dashed line.  The narrow, dark-gray shaded area represents the uncertainty in the relationship.  The light-gray shaded area illustrates deviations by a factor of 5 above and below the relation.  The location of J0330-2730 on this diagram is denoted by the magenta star.  This source falls significantly below the turnover-size relation established for PS sources, as described in Section~\ref{sec:turnover-size}.
\\
}
\label{fig:turnover_size}
\end{figure*}

CSOs are characterized by their small sizes ($<$1~kpc) and symmetric double-lobed morphologies \citep{odea+21}.  They are believed to represent an early stage in the evolution of radio-loud AGN. 
The resolved morphology of J0330-2730 in our VLBA images and its linear size of 64~pc is consistent with typical CSO definitions in the literature, although we note that a core has not yet been identified as required by some previous definitions (e.g. \citealt{orienti+14}). Higher-resolution VLBA imaging (i.e. at X-band) capable of isolating the position of a compact, self-absorbed core (or placing a firm upper limit on its presence) would help to clarify the morphological classification of this source.  

Recent studies of CSOs have focused on reducing  contamination by blazars that affects literature CSO samples.  
\citet{kiehlmann+24} proposed two constraints on the defining criteria for CSOs in addition to their compact, symmetrical morphologies to distinguish them from blazars: 1) variability and 2) superluminal motion.  The \citet{kiehlmann+24} criteria require CSOs to have superluminal motion $<$2.5c and variability $<20\%$. 
J0330-2730 satisfies the variability criterion based on constraints from times series data from both VLITE (340~MHz) and VLASS (3~GHz).  

The VLBA observations presented in this paper are the only observations of J0330-2730 so far with milliarcsecond-scale resolution.  Thus, no information on the jet expansion speed from proper motion measurements is available.       
Additional observations of J0330-2730 with the VLBA over the next few years will be needed to measure its expansion speed and definitively rule-out the presence of any superluminal motion. 

Overall, we conclude that our source is consistent with being a member of the CSO class, but we caution that additional VLBA observations are warranted.  
We suggest that future observations should focus on measuring the location of the core, determining the frequency dependence of all compact source components, and monitoring the evolution of the source over time.

\subsection{Implications for Jet Evolution}

\subsubsection{Jet Triggering Mechanism}
\label{sec:jet_formation}
Understanding the physics of the formation of radio jets and the factors that influence their life cycles is essential for incorporating radio AGN into models of galaxy evolution \citep{tadhunter+16}.  The formation of jets is known to be closely linked to the properties of the SMBH and its accretion rate and state \citep{blandford+19}.  Large-scale radio AGN, such as classical FRI/FRII radio galaxies \citep{fanaroff+74} with sizes that extend well beyond their host galaxies, form and grow over long timescales (millions of years).  However, recent statistical analyses of the number of compact radio AGN, including CSOs, have argued the majority of these sources must be short-lived in order to explain their high number density in surveys \citep{czerny+09, gugliucci+05, an+12, nyland+20}.  The formation of short-lived radio jets may be associated with accretion disk instabilities or transient phenomena (e.g. \citealt{Kunert-Bajraszewska+24}).  

One transient phenomenon capable of triggering jet formation is a tidal disruption event (TDE; \citealt{hills+75}).  TDEs have recently been proposed as an important mechanism for the formation of short-lived CSOs \citep{readhead+24}.  
A TDE occurs when a star passes within the tidal radius of the SMBH and is gravitationally shredded \citep{komossa+15}.  TDEs of solar-mass stars by non-spinning SMBHs do not occur above a mass of 10$^8$ M$_{\odot}$ \citep{stone+16}.  However, a solar-mass TDE associated with a maximally spinning SMBH \citep{kesden+12}, the disruption of an evolved star with a larger radius \citep{macleod+12}, or more exotic scenarios \citep{ryu+24}, remains plausible.  

Given the uncertain nature of the SMBH mass estimate for J0330-2730 presented in this study of 10$^8$ M$_{\odot}$, 
we cannot draw any firm conclusions on the feasibility of a TDE-like origin for the compact radio source.  A more robust SMBH mass estimate, as well as spectroscopic studies to measure the age of the stellar population in the vicinity of the AGN, would provide additional constraints.  
However, we note that the 3~GHz luminosity of J0330-2730 of log($L_{3\,{\rm GHz}}$/W Hz$^{-1}$) = 26.03 is more than an order of magnitude brighter than the most luminous TDEs known \citep{cendes+24}.  

We caution that a more precise SMBH mass estimate (e.g. based on spectroscopy) will be needed to further refine the properties of the AGN, constrain the possibility of a TDE origin for the jet triggering, and to predict the 
future evolution of the radio source.  Additional VLBA observations to better characterize the morphology, spectral index, variability, proper motion, and polarimetric properties of this source will provide further insight into its absorption physics, environment, triggering, and evolution.


\subsubsection{Turnover-size Relation}
\label{sec:turnover-size}
To further explore the evolutionary stage of the source, we consider the turnover-size relation.  
The turnover-size relation is an empirical anti-correlation between the peak or turnover frequency of the radio spectrum and the linear extent of the source \citep{odea+97}.  This relationship has been shown to agree well with models based on SSA, self-similar source expansion, and equipartition \citep{snellen+00, devries+09, jeyakumar+16}.  While alternative mechanisms, such as FFA, may offer a plausible explanation for some individual sources \citep{bicknell+97, bicknell+18}, the turnover-size relation for samples spanning a wide range of parameters values is best described by SSA \citep{odea+21}.

In Figure~\ref{fig:turnover_size}, we show the turnover-size relation for a sample of peaked-spectrum sources from the literature.  
J0330-2730 has a rest-frame turnover frequency of 582~MHz, which implies a size of $\sim$1~kpc ($\approx$ 0.3$^{\prime \prime}$).  This is a factor of $\sim$16 times larger
than the resolved extent measured with the VLBA of 64~pc. 
J0330-2730 is therefore significantly smaller than expected. 
This finding is consistent with recent VLBI imaging studies of MPS sources that have found that a substantial fraction of this population is more compact than predicted by the turnover-size relation \citep{keim+19}.  

Possible explanations for the departure from the turnover-size relation include an intermittent jet fuel supply or a ``frustrated jet'' impeded by the presence of dense gas or propagating in a high-pressure environment.  We review these possibilities in detail in the remainder of this section.  
In Section~\ref{sec:jet_formation}, we discussed the possibility of a jet formation pathway related to a transient phenomenon such as a TDE.  We speculate that a TDE origin could conceivably cause the departure from the turnover size relation if TDE evolution deviates from that of young radio AGN. 
Another possibility is an intermittent fuel supply due to accretion disk instabilities.  Recent simulations by \citet{lalakos+24} have shown that unstable or intermittent accretion can lead to episodic, short-lived jets with properties similar to some populations of compact radio AGN (e.g. FR0 source; \citealt{baldi+18}). 

The most popular explanation in the literature for sources that fall below the turnover-size relation is the frustrated jet scenario (e.g. \citealt{ballieux+24}).  Frustrated jets may represent a population of radio AGN that do not undergo the typical evolution into 
FRI/FRII sources.  In this model, the radio source's expansion is significantly impeded by interaction with a dense and/or clumpy ISM, causing it to remain more compact than expected for its age and turnover frequency. Numerous studies have found evidence that CSOs reside in dense environments that could influence their propagation (e.g. \citealt{kosmaczewski+20}).  This explanation is also supported by recent simulations demonstrating the effects of an inhomogenous, clumpy medium on jet evolution \citep{mukherjee+18}.  

In addition to the compact source size, other possible observational signatures of jet frustration or deceleration include a complex, disturbed morphology (e.g. \citealt{zovaro+19}), low-frequency variability due to FFA \citep{tingay+15}, 
gas outflows \citep{holt+09, morganti+13}, and Faraday rotation \citep{pasetto+16}.  The 340~MHz variability of J0330-2730 discussed in Section~\ref{sec:ffa} may be consistent with the jet frustration scenario, but its relatively symmetric morphology revealed by our C-band VLBA image suggests other explanations should be considered.  
While frustrated jets are commonly viewed as being associated with dense gas and direct jet-ISM interactions, pressure confinement can also lead to frustration \citep{an+12, sobolewska+19, perucho+17}.  

We conclude that the underlying cause of the departure from the turnover-size relation remains uncertain for this source.  Future observations of the pc-scale radio emission are needed to determine the polarization properties of the source, the spectral indices of the lobes, and the gas content and conditions in the ambient environment of the jet.  These measurements will help us better understand the relationship between the formation and evolution of J0330-2730 and its environment.

\section{Summary}
\label{sec:summary}
We have presented new VLBA observations of the previously unknown MPS source J0330-2730 that was identified using commensal 340~MHz VLITE data.  Our main conclusions are as follows: 

\begin{itemize}

   \item  We performed radio spectral modeling by combining new, archival, and survey flux measurements spanning 150~MHz to 20~GHz.  The source has a peaked spectral shape that is well described by either a SSA or FFA model with a spectral turnover frequency of 307~MHz (582~MHz in the rest frame of the source at its photometric redshift of $z  =  0.8975$).

    \item We found moderate but statistically significant variability at 340~MHz in our VLITE measurements over two epochs spanning 1.3~years.  We considered three potential mechanisms that could explain the observed variability: refractive interstellar scintillation, variable free-free absorption, and relativistic beaming.  We ruled-out relativistic beaming and consider both RISS and variable FFA to be plausible explanations given the currently available radio data.  
    
    \item Our VLBA observations revealed a spatially-resolved source at C band that has a double-lobed morphology with a projected linear size of 64~pc. 
    Based on the mild flux asymmetry of the lobes, we estimate a jet orientation angle of 75$^{\circ}$.  
       
    \item The size of J0330-2730's jets/lobes is significantly smaller (by a factor of $\sim$16) than the value predicted from the turnover-size relation. We conclude that the reason for the departure from the turnover-size relation is uncertain, but may be related to jet frustration or pressure confinement.  We also note that a transient origin for the formation of the jet, such as a TDE, cannot be ruled out.  
    
    \item We analyzed the radio variability of CSOs and blazars at VLITE’s 340 MHz frequency using the maximum fractional variability metric from \citet{kiehlmann+24b}, originally applied at GHz frequencies. This metric also appears effective at 340 MHz, with the majority of CSOs exhibiting $<20$\% yr$^{-1}$ variability. 
    However, we note that the small CSO sample size introduces some uncertainty in this result.

    \item Based on its relatively symmetric morphology, 
    moderate level of radio variability, and constraints on variable Doppler boosting, we conclude that relativistic beaming is not significant for J0330-2730.  The radio properties of this source are most consistent with it being a CSO.  

    \item We performed SED modeling and showed that the host galaxy is consistent with a massive galaxy harboring a low-luminosity AGN.  We roughly estimate a SMBH mass of $\sim10^8$ M$_{\odot}$.  

    \item This study highlights the potential of using low-frequency commensal radio instruments like VLITE to aid in the identification of compact and potentially young radio sources such as CSOs.
\end{itemize}

We emphasize that determining the physical mechanisms that influence radio jet lifetimes and triggering timescales is crucial for improving our understanding of jet formation, growth, and evolution, as well as for quantifying the impact of jet-driven feedback on galaxy evolution.  Additional observations with the VLBA, and future radio telescopes such as the next-generation Very Large Array \citep{murphy+18, nyland+18} and the Square Kilometre Array \citep{dewdney+09},  will ultimately be needed to gain further insight into the connection between the evolution of AGN jets and their host galaxies. 


\begin{acknowledgments}
We thank the anonymous referee for providing thoughtful feedback that has improved this manuscript.  We thank Dipanjan Mukherjee for helpful comments that have improved this manuscript.  The National Radio Astronomy Observatory is a facility of the National Science Foundation operated under cooperative agreement by Associated Universities, Inc.  
Basic research in radio astronomy at the U.S. Naval Research Laboratory is supported by 6.1 Base Funding. Construction and installation of VLITE was supported by the NRL Sustainment Restoration and Maintenance fund.  This work is based in part on observations made with the {\it Spitzer} Space Telescope, which was operated by the Jet Propulsion Laboratory, California Institute of Technology under a contract with NASA.  This research has made use of the NASA/IPAC Infrared Science Archive, which is funded by the National Aeronautics and Space Administration and operated by the California Institute of Technology.
\end{acknowledgments}

\vspace{5mm}
\facilities{VLA, VLBA} 

\software{Astropy \citep{astropy+13},  
          CASA \citep{casa+22}, AIPS \citep{greisen+03}, Obit \citep{cotton+08}, PyBDSF \citep{mohan+15}
          }



\bibliography{main}{}

\begin{thebibliography}{}
\expandafter\ifx\csname natexlab\endcsname\relax\def\natexlab#1{#1}\fi
\providecommand{\url}[1]{\href{#1}{#1}}
\providecommand{\dodoi}[1]{doi:~\href{http://doi.org/#1}{\nolinkurl{#1}}}
\providecommand{\doeprint}[1]{\href{http://ascl.net/#1}{\nolinkurl{http://ascl.net/#1}}}
\providecommand{\doarXiv}[1]{\href{https://arxiv.org/abs/#1}{\nolinkurl{https://arxiv.org/abs/#1}}}

\bibitem[{{Alatalo} {et~al.}(2011){Alatalo}, {Blitz}, {Young}, {Davis}, {Bureau}, {Lopez}, {Cappellari}, {Scott}, {Shapiro}, {Crocker}, {Mart{\'\i}n}, {Bois}, {Bournaud}, {Davies}, {de Zeeuw}, {Duc}, {Emsellem}, {Falc{\'o}n-Barroso}, {Khochfar}, {Krajnovi{\'c}}, {Kuntschner}, {Lablanche}, {McDermid}, {Morganti}, {Naab}, {Oosterloo}, {Sarzi}, {Serra}, \& {Weijmans}}]{alatalo+11}
{Alatalo}, K., {Blitz}, L., {Young}, L.~M., {et~al.} 2011, \apj, 735, 88, \dodoi{10.1088/0004-637X/735/2/88}

\bibitem[{{Alberts} {et~al.}(2020){Alberts}, {Rujopakarn}, {Rieke}, {Jagannathan}, \& {Nyland}}]{alberts+20}
{Alberts}, S., {Rujopakarn}, W., {Rieke}, G.~H., {Jagannathan}, P., \& {Nyland}, K. 2020, \apj, 901, 168, \dodoi{10.3847/1538-4357/abb1a0}

\bibitem[{{An} \& {Baan}(2012)}]{an+12}
{An}, T., \& {Baan}, W.~A. 2012, \apj, 760, 77, \dodoi{10.1088/0004-637X/760/1/77}

\bibitem[{{Antonucci}(1993)}]{antonucci+93}
{Antonucci}, R. 1993, \araa, 31, 473, \dodoi{10.1146/annurev.aa.31.090193.002353}

\bibitem[{{Arshakian} \& {Longair}(2004)}]{arshakian+04}
{Arshakian}, T.~G., \& {Longair}, M.~S. 2004, \mnras, 351, 727, \dodoi{10.1111/j.1365-2966.2004.07823.x}

\bibitem[{{Astropy Collaboration} {et~al.}(2013){Astropy Collaboration}, {Robitaille}, {Tollerud}, {Greenfield}, {Droettboom}, {Bray}, {Aldcroft}, {Davis}, {Ginsburg}, {Price-Whelan}, {Kerzendorf}, {Conley}, {Crighton}, {Barbary}, {Muna}, {Ferguson}, {Grollier}, {Parikh}, {Nair}, {Unther}, {Deil}, {Woillez}, {Conseil}, {Kramer}, {Turner}, {Singer}, {Fox}, {Weaver}, {Zabalza}, {Edwards}, {Azalee Bostroem}, {Burke}, {Casey}, {Crawford}, {Dencheva}, {Ely}, {Jenness}, {Labrie}, {Lim}, {Pierfederici}, {Pontzen}, {Ptak}, {Refsdal}, {Servillat}, \& {Streicher}}]{astropy+13}
{Astropy Collaboration}, {Robitaille}, T.~P., {Tollerud}, E.~J., {et~al.} 2013, \aap, 558, A33, \dodoi{10.1051/0004-6361/201322068}

\bibitem[{{Baldi} {et~al.}(2018){Baldi}, {Capetti}, \& {Massaro}}]{baldi+18}
{Baldi}, R.~D., {Capetti}, A., \& {Massaro}, F. 2018, \aap, 609, A1, \dodoi{10.1051/0004-6361/201731333}

\bibitem[{{Ballieux} {et~al.}(2024){Ballieux}, {Callingham}, {R{\"o}ttgering}, \& {Slob}}]{ballieux+24}
{Ballieux}, F.~J., {Callingham}, J.~R., {R{\"o}ttgering}, H.~J.~A., \& {Slob}, M.~M. 2024, \aap, 689, A264, \dodoi{10.1051/0004-6361/202449675}

\bibitem[{{Barvainis} {et~al.}(2005){Barvainis}, {Leh{\'a}r}, {Birkinshaw}, {Falcke}, \& {Blundell}}]{barvainis+05}
{Barvainis}, R., {Leh{\'a}r}, J., {Birkinshaw}, M., {Falcke}, H., \& {Blundell}, K.~M. 2005, \apj, 618, 108, \dodoi{10.1086/425859}

\bibitem[{{Beckwith} {et~al.}(2006){Beckwith}, {Stiavelli}, {Koekemoer}, {Caldwell}, {Ferguson}, {Hook}, {Lucas}, {Bergeron}, {Corbin}, {Jogee}, {Panagia}, {Robberto}, {Royle}, {Somerville}, \& {Sosey}}]{beckwith+06}
{Beckwith}, S. V.~W., {Stiavelli}, M., {Koekemoer}, A.~M., {et~al.} 2006, \aj, 132, 1729, \dodoi{10.1086/507302}

\bibitem[{{Bell} {et~al.}(2019){Bell}, {Murphy}, {Hancock}, {Callingham}, {Johnston}, {Kaplan}, {Hunstead}, {Sadler}, {Croft}, {White}, {Hurley-Walker}, {Chhetri}, {Morgan}, {Edwards}, {Rowlinson}, {Offringa}, {Bernardi}, {Bowman}, {Briggs}, {Cappallo}, {Deshpande}, {Gaensler}, {Greenhill}, {Hazelton}, {Johnston-Hollitt}, {Lonsdale}, {McWhirter}, {Mitchell}, {Morales}, {Morgan}, {Oberoi}, {Ord}, {Prabu}, {Shankar}, {Srivani}, {Subrahmanyan}, {Tingay}, {Wayth}, {Webster}, {Williams}, \& {Williams}}]{bell+19}
{Bell}, M.~E., {Murphy}, T., {Hancock}, P.~J., {et~al.} 2019, \mnras, 482, 2484, \dodoi{10.1093/mnras/sty2801}

\bibitem[{{Best} {et~al.}(2005){Best}, {Kauffmann}, {Heckman}, \& {Ivezi{\'c}}}]{best+05}
{Best}, P.~N., {Kauffmann}, G., {Heckman}, T.~M., \& {Ivezi{\'c}}, {\v Z}. 2005, \mnras, 362, 9, \dodoi{10.1111/j.1365-2966.2005.09283.x}

\bibitem[{{Bicknell} {et~al.}(1997){Bicknell}, {Dopita}, \& {O'Dea}}]{bicknell+97}
{Bicknell}, G.~V., {Dopita}, M.~A., \& {O'Dea}, C. P.~O. 1997, \apj, 485, 112, \dodoi{10.1086/304400}

\bibitem[{{Bicknell} {et~al.}(2018){Bicknell}, {Mukherjee}, {Wagner}, {Sutherland}, \& {Nesvadba}}]{bicknell+18}
{Bicknell}, G.~V., {Mukherjee}, D., {Wagner}, A. e.~Y., {Sutherland}, R.~S., \& {Nesvadba}, N. P.~H. 2018, \mnras, 475, 3493, \dodoi{10.1093/mnras/sty070}

\bibitem[{{Blandford} {et~al.}(2019){Blandford}, {Meier}, \& {Readhead}}]{blandford+19}
{Blandford}, R., {Meier}, D., \& {Readhead}, A. 2019, \araa, 57, 467, \dodoi{10.1146/annurev-astro-081817-051948}

\bibitem[{{Boquien} {et~al.}(2019){Boquien}, {Burgarella}, {Roehlly}, {Buat}, {Ciesla}, {Corre}, {Inoue}, \& {Salas}}]{boquien+19}
{Boquien}, M., {Burgarella}, D., {Roehlly}, Y., {et~al.} 2019, \aap, 622, A103, \dodoi{10.1051/0004-6361/201834156}

\bibitem[{{Bruzual} \& {Charlot}(2003)}]{bc03}
{Bruzual}, G., \& {Charlot}, S. 2003, \mnras, 344, 1000, \dodoi{10.1046/j.1365-8711.2003.06897.x}

\bibitem[{{Callingham} {et~al.}(2017){Callingham}, {Ekers}, {Gaensler}, {Line}, {Hurley-Walker}, {Sadler}, {Tingay}, {Hancock}, {Bell}, {Dwarakanath}, {For}, {Franzen}, {Hindson}, {Johnston-Hollitt}, {Kapi{\'n}ska}, {Lenc}, {McKinley}, {Morgan}, {Offringa}, {Procopio}, {Staveley-Smith}, {Wayth}, {Wu}, \& {Zheng}}]{callingham+17}
{Callingham}, J.~R., {Ekers}, R.~D., {Gaensler}, B.~M., {et~al.} 2017, \apj, 836, 174, \dodoi{10.3847/1538-4357/836/2/174}

\bibitem[{{CASA Team} {et~al.}(2022){CASA Team}, {Bean}, {Bhatnagar}, {Castro}, {Donovan Meyer}, {Emonts}, {Garcia}, {Garwood}, {Golap}, {Gonzalez Villalba}, {Harris}, {Hayashi}, {Hoskins}, {Hsieh}, {Jagannathan}, {Kawasaki}, {Keimpema}, {Kettenis}, {Lopez}, {Marvil}, {Masters}, {McNichols}, {Mehringer}, {Miel}, {Moellenbrock}, {Montesino}, {Nakazato}, {Ott}, {Petry}, {Pokorny}, {Raba}, {Rau}, {Schiebel}, {Schweighart}, {Sekhar}, {Shimada}, {Small}, {Steeb}, {Sugimoto}, {Suoranta}, {Tsutsumi}, {van Bemmel}, {Verkouter}, {Wells}, {Xiong}, {Szomoru}, {Griffith}, {Glendenning}, \& {Kern}}]{casa+22}
{CASA Team}, {Bean}, B., {Bhatnagar}, S., {et~al.} 2022, \pasp, 134, 114501, \dodoi{10.1088/1538-3873/ac9642}

\bibitem[{{Cendes} {et~al.}(2024){Cendes}, {Berger}, {Alexander}, {Chornock}, {Margutti}, {Metzger}, {Wieringa}, {Bietenholz}, {Hajela}, {Laskar}, {Stroh}, \& {Terreran}}]{cendes+24}
{Cendes}, Y., {Berger}, E., {Alexander}, K.~D., {et~al.} 2024, \apj, 971, 185, \dodoi{10.3847/1538-4357/ad5541}

\bibitem[{{Clarke} {et~al.}(2016){Clarke}, {Kassim}, {Brisken}, {Helmboldt}, {Peters}, {Ray}, {Polisensky}, \& {Giacintucci}}]{clarke+16}
{Clarke}, T.~E., {Kassim}, N.~E., {Brisken}, W., {et~al.} 2016, in Society of Photo-Optical Instrumentation Engineers (SPIE) Conference Series, Vol. 9906, Ground-based and Airborne Telescopes VI, 99065B, \dodoi{10.1117/12.2233036}

\bibitem[{{Condon} {et~al.}(1998){Condon}, {Cotton}, {Greisen}, {Yin}, {Perley}, {Taylor}, \& {Broderick}}]{condon+98}
{Condon}, J.~J., {Cotton}, W.~D., {Greisen}, E.~W., {et~al.} 1998, \aj, 115, 1693, \dodoi{10.1086/300337}

\bibitem[{{Condon} \& {Matthews}(2018)}]{condon+18}
{Condon}, J.~J., \& {Matthews}, A.~M. 2018, \pasp, 130, 073001, \dodoi{10.1088/1538-3873/aac1b2}

\bibitem[{{Cotton}(2008)}]{cotton+08}
{Cotton}, W.~D. 2008, \pasp, 120, 439, \dodoi{10.1086/586754}

\bibitem[{{Czerny} {et~al.}(2009){Czerny}, {Siemiginowska}, {Janiuk}, {Nikiel-Wroczy{\'n}ski}, \& {Stawarz}}]{czerny+09}
{Czerny}, B., {Siemiginowska}, A., {Janiuk}, A., {Nikiel-Wroczy{\'n}ski}, B., \& {Stawarz}, {\L}. 2009, \apj, 698, 840, \dodoi{10.1088/0004-637X/698/1/840}

\bibitem[{{de Vries} {et~al.}(2009){de Vries}, {Snellen}, {Schilizzi}, {Mack}, \& {Kaiser}}]{devries+09}
{de Vries}, N., {Snellen}, I.~A.~G., {Schilizzi}, R.~T., {Mack}, K.~H., \& {Kaiser}, C.~R. 2009, \aap, 498, 641, \dodoi{10.1051/0004-6361/200811145}

\bibitem[{{Dewdney} {et~al.}(2009){Dewdney}, {Hall}, {Schilizzi}, \& {Lazio}}]{dewdney+09}
{Dewdney}, P.~E., {Hall}, P.~J., {Schilizzi}, R.~T., \& {Lazio}, T. J. L.~W. 2009, Proceedings of the IEEE, 97, 1482

\bibitem[{{Dey} {et~al.}(2019){Dey}, {Schlegel}, {Lang}, {Blum}, {Burleigh}, {Fan}, {Findlay}, {Finkbeiner}, {Herrera}, {Juneau}, {Landriau}, {Levi}, {McGreer}, {Meisner}, {Myers}, {Moustakas}, {Nugent}, {Patej}, {Schlafly}, {Walker}, {Valdes}, {Weaver}, {Y{\`e}che}, {Zou}, {Zhou}, {Abareshi}, {Abbott}, {Abolfathi}, {Aguilera}, {Alam}, {Allen}, {Alvarez}, {Annis}, {Ansarinejad}, {Aubert}, {Beechert}, {Bell}, {BenZvi}, {Beutler}, {Bielby}, {Bolton}, {Brice{\~n}o}, {Buckley-Geer}, {Butler}, {Calamida}, {Carlberg}, {Carter}, {Casas}, {Castander}, {Choi}, {Comparat}, {Cukanovaite}, {Delubac}, {DeVries}, {Dey}, {Dhungana}, {Dickinson}, {Ding}, {Donaldson}, {Duan}, {Duckworth}, {Eftekharzadeh}, {Eisenstein}, {Etourneau}, {Fagrelius}, {Farihi}, {Fitzpatrick}, {Font-Ribera}, {Fulmer}, {G{\"a}nsicke}, {Gaztanaga}, {George}, {Gerdes}, {Gontcho}, {Gorgoni}, {Green}, {Guy}, {Harmer}, {Hernandez}, {Honscheid}, {Huang}, {James}, {Jannuzi}, {Jiang}, {Joyce}, {Karcher}, {Karkar}, {Kehoe}, {Kneib}, {Kueter-Young}, {Lan},
  {Lauer}, {Le Guillou}, {Le Van Suu}, {Lee}, {Lesser}, {Perreault Levasseur}, {Li}, {Mann}, {Marshall}, {Mart{\'\i}nez-V{\'a}zquez}, {Martini}, {du Mas des Bourboux}, {McManus}, {Meier}, {M{\'e}nard}, {Metcalfe}, {Mu{\~n}oz-Guti{\'e}rrez}, {Najita}, {Napier}, {Narayan}, {Newman}, {Nie}, {Nord}, {Norman}, {Olsen}, {Paat}, {Palanque-Delabrouille}, {Peng}, {Poppett}, {Poremba}, {Prakash}, {Rabinowitz}, {Raichoor}, {Rezaie}, {Robertson}, {Roe}, {Ross}, {Ross}, {Rudnick}, {Safonova}, {Saha}, {S{\'a}nchez}, {Savary}, {Schweiker}, {Scott}, {Seo}, {Shan}, {Silva}, {Slepian}, {Soto}, {Sprayberry}, {Staten}, {Stillman}, {Stupak}, {Summers}, {Sien Tie}, {Tirado}, {Vargas-Maga{\~n}a}, {Vivas}, {Wechsler}, {Williams}, {Yang}, {Yang}, {Yapici}, {Zaritsky}, {Zenteno}, {Zhang}, {Zhang}, {Zhou}, \& {Zhou}}]{dey+19}
{Dey}, A., {Schlegel}, D.~J., {Lang}, D., {et~al.} 2019, \aj, 157, 168, \dodoi{10.3847/1538-3881/ab089d}

\bibitem[{{Duchesne} {et~al.}(2024){Duchesne}, {Grundy}, {Heald}, {Lenc}, {Leung}, {McConnell}, {Murphy}, {Pritchard}, {Rose}, {Thomson}, {Wang}, {Wang}, \& {Whiting}}]{duchesne+24}
{Duchesne}, S.~W., {Grundy}, J.~A., {Heald}, G.~H., {et~al.} 2024, \pasa, 41, e003, \dodoi{10.1017/pasa.2023.60}

\bibitem[{{Fabian}(2012)}]{fabian+12}
{Fabian}, A.~C. 2012, \araa, 50, 455, \dodoi{10.1146/annurev-astro-081811-125521}

\bibitem[{{Fanaroff} \& {Riley}(1974)}]{fanaroff+74}
{Fanaroff}, B.~L., \& {Riley}, J.~M. 1974, \mnras, 167, 31P, \dodoi{10.1093/mnras/167.1.31P}

\bibitem[{{Ferrarese} \& {Merritt}(2000)}]{ferrarese+00}
{Ferrarese}, L., \& {Merritt}, D. 2000, \apjl, 539, L9, \dodoi{10.1086/312838}

\bibitem[{{Fiedler} {et~al.}(1987){Fiedler}, {Dennison}, {Johnston}, \& {Hewish}}]{fiedler+87}
{Fiedler}, R.~L., {Dennison}, B., {Johnston}, K.~J., \& {Hewish}, A. 1987, \nat, 326, 675, \dodoi{10.1038/326675a0}

\bibitem[{{Franzen} {et~al.}(2014){Franzen}, {Sadler}, {Chhetri}, {Ekers}, {Mahony}, {Murphy}, {Norris}, {Waldram}, \& {Whittam}}]{franzen+14}
{Franzen}, T. M.~O., {Sadler}, E.~M., {Chhetri}, R., {et~al.} 2014, \mnras, 439, 1212, \dodoi{10.1093/mnras/stt2322}

\bibitem[{{Franzen} {et~al.}(2015){Franzen}, {Banfield}, {Hales}, {Hopkins}, {Norris}, {Seymour}, {Chow}, {Herzog}, {Huynh}, {Lenc}, {Mao}, \& {Middelberg}}]{franzen+15}
{Franzen}, T.~M.~O., {Banfield}, J.~K., {Hales}, C.~A., {et~al.} 2015, \mnras, 453, 4020, \dodoi{10.1093/mnras/stv1866}

\bibitem[{{Gordon} {et~al.}(2021){Gordon}, {Boyce}, {O'Dea}, {Rudnick}, {Andernach}, {Vantyghem}, {Baum}, {Bui}, {Dionyssiou}, {Safi-Harb}, \& {Sander}}]{gordon+21}
{Gordon}, Y.~A., {Boyce}, M.~M., {O'Dea}, C.~P., {et~al.} 2021, \apjs, 255, 30, \dodoi{10.3847/1538-4365/ac05c0}

\bibitem[{{Greisen}(2003)}]{greisen+03}
{Greisen}, E.~W. 2003, in Astrophysics and Space Science Library, Vol. 285, Information Handling in Astronomy - Historical Vistas, ed. A.~{Heck}, 109, \dodoi{10.1007/0-306-48080-8_7}

\bibitem[{{Gugliucci} {et~al.}(2005){Gugliucci}, {Taylor}, {Peck}, \& {Giroletti}}]{gugliucci+05}
{Gugliucci}, N.~E., {Taylor}, G.~B., {Peck}, A.~B., \& {Giroletti}, M. 2005, \apj, 622, 136, \dodoi{10.1086/427934}

\bibitem[{{Hardcastle} \& {Croston}(2020)}]{hardcastle+20}
{Hardcastle}, M.~J., \& {Croston}, J.~H. 2020, \nar, 88, 101539, \dodoi{10.1016/j.newar.2020.101539}

\bibitem[{{Heckman} \& {Best}(2014)}]{heckman+14}
{Heckman}, T.~M., \& {Best}, P.~N. 2014, \araa, 52, 589, \dodoi{10.1146/annurev-astro-081913-035722}

\bibitem[{{Hills}(1975)}]{hills+75}
{Hills}, J.~G. 1975, \nat, 254, 295, \dodoi{10.1038/254295a0}

\bibitem[{{Ho}(2002)}]{ho+02}
{Ho}, L.~C. 2002, \apj, 564, 120, \dodoi{10.1086/324399}

\bibitem[{{Ho}(2008)}]{ho+08}
---. 2008, \araa, 46, 475, \dodoi{10.1146/annurev.astro.45.051806.110546}

\bibitem[{{Holt}(2009)}]{holt+09}
{Holt}, J. 2009, Astronomische Nachrichten, 330, 226, \dodoi{10.1002/asna.200811163}

\bibitem[{{Hovatta} \& {Lindfors}(2019)}]{hovatta+19}
{Hovatta}, T., \& {Lindfors}, E. 2019, \nar, 87, 101541, \dodoi{10.1016/j.newar.2020.101541}

\bibitem[{{Hunt} {et~al.}(2021){Hunt}, {Johnson}, {Cigan}, {Gordon}, \& {Spitzak}}]{hunt+21}
{Hunt}, L.~R., {Johnson}, M.~C., {Cigan}, P.~J., {Gordon}, D., \& {Spitzak}, J. 2021, \aj, 162, 121, \dodoi{10.3847/1538-3881/ac135d}

\bibitem[{{Hurley-Walker} {et~al.}(2017){Hurley-Walker}, {Callingham}, {Hancock}, {Franzen}, {Hindson}, {Kapi{\'n}ska}, {Morgan}, {Offringa}, {Wayth}, {Wu}, {Zheng}, {Murphy}, {Bell}, {Dwarakanath}, {For}, {Gaensler}, {Johnston-Hollitt}, {Lenc}, {Procopio}, {Staveley-Smith}, {Ekers}, {Bowman}, {Briggs}, {Cappallo}, {Deshpande}, {Greenhill}, {Hazelton}, {Kaplan}, {Lonsdale}, {McWhirter}, {Mitchell}, {Morales}, {Morgan}, {Oberoi}, {Ord}, {Prabu}, {Shankar}, {Srivani}, {Subrahmanyan}, {Tingay}, {Webster}, {Williams}, \& {Williams}}]{hurley-walker+17}
{Hurley-Walker}, N., {Callingham}, J.~R., {Hancock}, P.~J., {et~al.} 2017, \mnras, 464, 1146, \dodoi{10.1093/mnras/stw2337}

\bibitem[{{Intema} {et~al.}(2017){Intema}, {Jagannathan}, {Mooley}, \& {Frail}}]{intema+17}
{Intema}, H.~T., {Jagannathan}, P., {Mooley}, K.~P., \& {Frail}, D.~A. 2017, \aap, 598, A78, \dodoi{10.1051/0004-6361/201628536}

\bibitem[{{IRSA} \& {SSC}(2020)}]{seip+20}
{IRSA}, \& {SSC}. 2020, {Spitzer Enhanced Imaging Products}, NASA IPAC DataSet, IRSA433, \dodoi{10.26131/IRSA433}

\bibitem[{{Ivezi{\'c}} {et~al.}(2002){Ivezi{\'c}}, {Menou}, {Knapp}, {Strauss}, {Lupton}, {Vanden Berk}, {Richards}, {Tremonti}, {Weinstein}, {Anderson}, {Bahcall}, {Becker}, {Bernardi}, {Blanton}, {Eisenstein}, {Fan}, {Finkbeiner}, {Finlator}, {Frieman}, {Gunn}, {Hall}, {Kim}, {Kinkhabwala}, {Narayanan}, {Rockosi}, {Schlegel}, {Schneider}, {Strateva}, {SubbaRao}, {Thakar}, {Voges}, {White}, {Yanny}, {Brinkmann}, {Doi}, {Fukugita}, {Hennessy}, {Munn}, {Nichol}, \& {York}}]{ivezic+02}
{Ivezi{\'c}}, {\v{Z}}., {Menou}, K., {Knapp}, G.~R., {et~al.} 2002, \aj, 124, 2364, \dodoi{10.1086/344069}

\bibitem[{{Jarvis} {et~al.}(2013){Jarvis}, {Bonfield}, {Bruce}, {Geach}, {McAlpine}, {McLure}, {Gonz{\'a}lez-Solares}, {Irwin}, {Lewis}, {Yoldas}, {Andreon}, {Cross}, {Emerson}, {Dalton}, {Dunlop}, {Hodgkin}, {Le}, {Karouzos}, {Meisenheimer}, {Oliver}, {Rawlings}, {Simpson}, {Smail}, {Smith}, {Sullivan}, {Sutherland}, {White}, \& {Zwart}}]{jarvis+13}
{Jarvis}, M.~J., {Bonfield}, D.~G., {Bruce}, V.~A., {et~al.} 2013, \mnras, 428, 1281, \dodoi{10.1093/mnras/sts118}

\bibitem[{{Jeyakumar}(2016)}]{jeyakumar+16}
{Jeyakumar}, S. 2016, \mnras, 458, 3786, \dodoi{10.1093/mnras/stw181}

\bibitem[{{Keim} {et~al.}(2019){Keim}, {Callingham}, \& {R{\"o}ttgering}}]{keim+19}
{Keim}, M.~A., {Callingham}, J.~R., \& {R{\"o}ttgering}, H.~J.~A. 2019, \aap, 628, A56, \dodoi{10.1051/0004-6361/201936107}

\bibitem[{{Kellermann} \& {Pauliny-Toth}(1981)}]{kellerman+81}
{Kellermann}, K.~I., \& {Pauliny-Toth}, I.~I.~K. 1981, \araa, 19, 373, \dodoi{10.1146/annurev.aa.19.090181.002105}

\bibitem[{{Kesden}(2012)}]{kesden+12}
{Kesden}, M. 2012, \prd, 85, 024037, \dodoi{10.1103/PhysRevD.85.024037}

\bibitem[{{Kiehlmann} {et~al.}(2024{\natexlab{a}}){Kiehlmann}, {Readhead}, {O'Neill}, {Wilkinson}, {Lister}, {Liodakis}, {Bruzewski}, {Pavlidou}, {Pearson}, {Sheldahl}, {Siemiginowska}, {Tassis}, \& {Taylor}}]{kiehlmann+24b}
{Kiehlmann}, S., {Readhead}, A.~C.~S., {O'Neill}, S., {et~al.} 2024{\natexlab{a}}, \apj, 961, 241, \dodoi{10.3847/1538-4357/ad0cc2}

\bibitem[{{Kiehlmann} {et~al.}(2024{\natexlab{b}}){Kiehlmann}, {Lister}, {Readhead}, {Liodakis}, {O'Neill}, {Pearson}, {Sheldahl}, {Siemiginowska}, {Tassis}, {Taylor}, \& {Wilkinson}}]{kiehlmann+24}
{Kiehlmann}, S., {Lister}, M.~L., {Readhead}, A.~C.~S., {et~al.} 2024{\natexlab{b}}, \apj, 961, 240, \dodoi{10.3847/1538-4357/ad0c56}

\bibitem[{{Komossa}(2015)}]{komossa+15}
{Komossa}, S. 2015, Journal of High Energy Astrophysics, 7, 148, \dodoi{10.1016/j.jheap.2015.04.006}

\bibitem[{{Kormendy} \& {Ho}(2013)}]{kormendy+13}
{Kormendy}, J., \& {Ho}, L.~C. 2013, \araa, 51, 511, \dodoi{10.1146/annurev-astro-082708-101811}

\bibitem[{{Kormendy} \& {Richstone}(1995)}]{kormendy+95}
{Kormendy}, J., \& {Richstone}, D. 1995, \araa, 33, 581, \dodoi{10.1146/annurev.aa.33.090195.003053}

\bibitem[{{Kosmaczewski} {et~al.}(2020){Kosmaczewski}, {Stawarz}, {Siemiginowska}, {Cheung}, {Ostorero}, {Sobolewska}, {Kozie{\l}-Wierzbowska}, {W{\'o}jtowicz}, \& {Marchenko}}]{kosmaczewski+20}
{Kosmaczewski}, E., {Stawarz}, {\L}., {Siemiginowska}, A., {et~al.} 2020, \apj, 897, 164, \dodoi{10.3847/1538-4357/ab9b1f}

\bibitem[{{Kukreti} \& {Morganti}(2024)}]{kukreti+24}
{Kukreti}, P., \& {Morganti}, R. 2024, \aap, 690, A140, \dodoi{10.1051/0004-6361/202450454}

\bibitem[{{Kunert-Bajraszewska} {et~al.}(2024){Kunert-Bajraszewska}, {Krauze}, {Kimball}, {Stawarz}, {Kharb}, {Stern}, {Mooley}, {Nyland}, \& {Koziel-Wierzbowska}}]{Kunert-Bajraszewska+24}
{Kunert-Bajraszewska}, M., {Krauze}, A., {Kimball}, A.~E., {et~al.} 2024, arXiv e-prints, arXiv:2412.07702, \dodoi{10.48550/arXiv.2412.07702}

\bibitem[{{Lacy} {et~al.}(2020){Lacy}, {Baum}, {Chandler}, {Chatterjee}, {Clarke}, {Deustua}, {English}, {Farnes}, {Gaensler}, {Gugliucci}, {Hallinan}, {Kent}, {Kimball}, {Law}, {Lazio}, {Marvil}, {Mao}, {Medlin}, {Mooley}, {Murphy}, {Myers}, {Osten}, {Richards}, {Rosolowsky}, {Rudnick}, {Schinzel}, {Sivakoff}, {Sjouwerman}, {Taylor}, {White}, {Wrobel}, {Andernach}, {Beasley}, {Berger}, {Bhatnager}, {Birkinshaw}, {Bower}, {Brandt}, {Brown}, {Burke-Spolaor}, {Butler}, {Comerford}, {Demorest}, {Fu}, {Giacintucci}, {Golap}, {G{\"u}th}, {Hales}, {Hiriart}, {Hodge}, {Horesh}, {Ivezi{\'c}}, {Jarvis}, {Kamble}, {Kassim}, {Liu}, {Loinard}, {Lyons}, {Masters}, {Mezcua}, {Moellenbrock}, {Mroczkowski}, {Nyland}, {O{\textquoteright}Dea}, {O{\textquoteright}Sullivan}, {Peters}, {Radford}, {Rao}, {Robnett}, {Salcido}, {Shen}, {Sobotka}, {Witz}, {Vaccari}, {van Weeren}, {Vargas}, {Williams}, \& {Yoon}}]{lacy+20}
{Lacy}, M., {Baum}, S.~A., {Chandler}, C.~J., {et~al.} 2020, \pasp, 132, 035001, \dodoi{10.1088/1538-3873/ab63eb}

\bibitem[{{Lacy} {et~al.}(2021){Lacy}, {Surace}, {Farrah}, {Nyland}, {Afonso}, {Brandt}, {Clements}, {Lagos}, {Maraston}, {Pforr}, {Sajina}, {Sako}, {Vaccari}, {Wilson}, {Ballantyne}, {Barkhouse}, {Brunner}, {Cane}, {Clarke}, {Cooper}, {Cooray}, {Covone}, {D'Andrea}, {Evrard}, {Ferguson}, {Frieman}, {Gonzalez-Perez}, {Gupta}, {Hatziminaoglou}, {Huang}, {Jagannathan}, {Jarvis}, {Jones}, {Kimball}, {Lidman}, {Lubin}, {Marchetti}, {Martini}, {McMahon}, {Mei}, {Messias}, {Murphy}, {Newman}, {Nichol}, {Norris}, {Oliver}, {Perez-Fournon}, {Peters}, {Pierre}, {Polisensky}, {Richards}, {Ridgway}, {R{\"o}ttgering}, {Seymour}, {Shirley}, {Somerville}, {Strauss}, {Suntzeff}, {Thorman}, {van Kampen}, {Verma}, {Wechsler}, \& {Wood-Vasey}}]{lacy+21}
{Lacy}, M., {Surace}, J.~A., {Farrah}, D., {et~al.} 2021, \mnras, 501, 892, \dodoi{10.1093/mnras/staa3714}

\bibitem[{{Lalakos} {et~al.}(2024){Lalakos}, {Tchekhovskoy}, {Bromberg}, {Gottlieb}, {Jacquemin-Ide}, {Liska}, \& {Zhang}}]{lalakos+24}
{Lalakos}, A., {Tchekhovskoy}, A., {Bromberg}, O., {et~al.} 2024, \apj, 964, 79, \dodoi{10.3847/1538-4357/ad0974}

\bibitem[{{Liodakis} {et~al.}(2017){Liodakis}, {Pavlidou}, {Hovatta}, {Max-Moerbeck}, {Pearson}, {Richards}, \& {Readhead}}]{liodakis+17}
{Liodakis}, I., {Pavlidou}, V., {Hovatta}, T., {et~al.} 2017, \mnras, 467, 4565, \dodoi{10.1093/mnras/stx432}

\bibitem[{{Lister} {et~al.}(2013){Lister}, {Aller}, {Aller}, {Homan}, {Kellermann}, {Kovalev}, {Pushkarev}, {Richards}, {Ros}, \& {Savolainen}}]{lister+13}
{Lister}, M.~L., {Aller}, M.~F., {Aller}, H.~D., {et~al.} 2013, \aj, 146, 120, \dodoi{10.1088/0004-6256/146/5/120}

\bibitem[{{Lyu} {et~al.}(2022){Lyu}, {Alberts}, {Rieke}, \& {Rujopakarn}}]{lyu+22}
{Lyu}, J., {Alberts}, S., {Rieke}, G.~H., \& {Rujopakarn}, W. 2022, \apj, 941, 191, \dodoi{10.3847/1538-4357/ac9e5d}

\bibitem[{{MacLeod} {et~al.}(2012){MacLeod}, {Guillochon}, \& {Ramirez-Ruiz}}]{macleod+12}
{MacLeod}, M., {Guillochon}, J., \& {Ramirez-Ruiz}, E. 2012, \apj, 757, 134, \dodoi{10.1088/0004-637X/757/2/134}

\bibitem[{{Magorrian} {et~al.}(1998){Magorrian}, {Tremaine}, {Richstone}, {Bender}, {Bower}, {Dressler}, {Faber}, {Gebhardt}, {Green}, {Grillmair}, {Kormendy}, \& {Lauer}}]{magorrian+98}
{Magorrian}, J., {Tremaine}, S., {Richstone}, D., {et~al.} 1998, \aj, 115, 2285, \dodoi{10.1086/300353}

\bibitem[{{Marscher} \& {Gear}(1985)}]{marscher+85}
{Marscher}, A.~P., \& {Gear}, W.~K. 1985, \apj, 298, 114, \dodoi{10.1086/163592}

\bibitem[{{Massaro} {et~al.}(2015){Massaro}, {Maselli}, {Leto}, {Marchegiani}, {Perri}, {Giommi}, \& {Piranomonte}}]{massaro+15}
{Massaro}, E., {Maselli}, A., {Leto}, C., {et~al.} 2015, \apss, 357, 75, \dodoi{10.1007/s10509-015-2254-2}

\bibitem[{{McConnell} {et~al.}(2020){McConnell}, {Hale}, {Lenc}, {Banfield}, {Heald}, {Hotan}, {Leung}, {Moss}, {Murphy}, {O'Brien}, {Pritchard}, {Raja}, {Sadler}, {Stewart}, {Thomson}, {Whiting}, {Allison}, {Amy}, {Anderson}, {Ball}, {Bannister}, {Bell}, {Bock}, {Bolton}, {Bunton}, {Chippendale}, {Collier}, {Cooray}, {Cornwell}, {Diamond}, {Edwards}, {Gupta}, {Hayman}, {Heywood}, {Jackson}, {Koribalski}, {Lee-Waddell}, {McClure-Griffiths}, {Ng}, {Norris}, {Phillips}, {Reynolds}, {Roxby}, {Schinckel}, {Shields}, {Tremblay}, {Tzioumis}, {Voronkov}, \& {Westmeier}}]{mcconnell+20}
{McConnell}, D., {Hale}, C.~L., {Lenc}, E., {et~al.} 2020, \pasa, 37, e048, \dodoi{10.1017/pasa.2020.41}

\bibitem[{{Miley}(1980)}]{miley+80}
{Miley}, G. 1980, \araa, 18, 165, \dodoi{10.1146/annurev.aa.18.090180.001121}

\bibitem[{{Mohan} \& {Rafferty}(2015)}]{mohan+15}
{Mohan}, N., \& {Rafferty}, D. 2015, {PyBDSF: Python Blob Detection and Source Finder}, Astrophysics Source Code Library, record ascl:1502.007

\bibitem[{{Mooney} {et~al.}(2019){Mooney}, {Quinn}, {Callingham}, {Morganti}, {Duncan}, {Morabito}, {Best}, {G{\"u}rkan}, {Hardcastle}, {Prandoni}, {R{\"o}ttgering}, {Sabater}, {Shimwell}, {Shulevski}, {Tasse}, \& {Williams}}]{mooney+19}
{Mooney}, S., {Quinn}, J., {Callingham}, J.~R., {et~al.} 2019, \aap, 622, A14, \dodoi{10.1051/0004-6361/201833937}

\bibitem[{{Morabito} {et~al.}(2025){Morabito}, {Jackson}, {de Jong}, {Escott}, {Groeneveld}, {Mahatma}, {Petley}, {Sweijen}, {Timmerman}, \& {van Weeren}}]{morabito+25}
{Morabito}, L.~K., {Jackson}, N., {de Jong}, J., {et~al.} 2025, arXiv e-prints, arXiv:2502.06946, \dodoi{10.48550/arXiv.2502.06946}

\bibitem[{{Morgan} {et~al.}(2018){Morgan}, {Macquart}, {Ekers}, {Chhetri}, {Tokumaru}, {Manoharan}, {Tremblay}, {Bisi}, \& {Jackson}}]{morgan+18}
{Morgan}, J.~S., {Macquart}, J.~P., {Ekers}, R., {et~al.} 2018, \mnras, 473, 2965, \dodoi{10.1093/mnras/stx2284}

\bibitem[{{Morganti}(2017)}]{morganti+17}
{Morganti}, R. 2017, Frontiers in Astronomy and Space Sciences, 4, 42, \dodoi{10.3389/fspas.2017.00042}

\bibitem[{{Morganti} {et~al.}(2013){Morganti}, {Fogasy}, {Paragi}, {Oosterloo}, \& {Orienti}}]{morganti+13}
{Morganti}, R., {Fogasy}, J., {Paragi}, Z., {Oosterloo}, T., \& {Orienti}, M. 2013, Science, 341, 1082, \dodoi{10.1126/science.1240436}

\bibitem[{{Mukherjee} {et~al.}(2016){Mukherjee}, {Bicknell}, {Sutherland }, \& {Wagner}}]{mukherjee+16}
{Mukherjee}, D., {Bicknell}, G.~V., {Sutherland }, R., \& {Wagner}, A. 2016, \mnras, 461, 967, \dodoi{10.1093/mnras/stw1368}

\bibitem[{{Mukherjee} {et~al.}(2018){Mukherjee}, {Bicknell}, {Wagner}, {Sutherland}, \& {Silk}}]{mukherjee+18}
{Mukherjee}, D., {Bicknell}, G.~V., {Wagner}, A.~Y., {Sutherland}, R.~S., \& {Silk}, J. 2018, \mnras, 479, 5544, \dodoi{10.1093/mnras/sty1776}

\bibitem[{{Murphy} {et~al.}(2018){Murphy}, {Bolatto}, {Chatterjee}, {Casey}, {Chomiuk}, {Dale}, {de Pater}, {Dickinson}, {Francesco}, {Hallinan}, {Isella}, {Kohno}, {Kulkarni}, {Lang}, {Lazio}, {Leroy}, {Loinard}, {Maccarone}, {Matthews}, {Osten}, {Reid}, {Riechers}, {Sakai}, {Walter}, \& {Wilner}}]{murphy+18}
{Murphy}, E.~J., {Bolatto}, A., {Chatterjee}, S., {et~al.} 2018, in Astronomical Society of the Pacific Conference Series, Vol. 517, Science with a Next Generation Very Large Array, ed. E.~{Murphy}, 3, \dodoi{10.48550/arXiv.1810.07524}

\bibitem[{{Murthy} {et~al.}(2024){Murthy}, {Morganti}, {Oosterloo}, {Schulz}, \& {Paragi}}]{murthy+24}
{Murthy}, S., {Morganti}, R., {Oosterloo}, T., {Schulz}, R., \& {Paragi}, Z. 2024, \aap, 688, A84, \dodoi{10.1051/0004-6361/202450233}

\bibitem[{{Netzer}(2015)}]{netzer+15}
{Netzer}, H. 2015, \araa, 53, 365, \dodoi{10.1146/annurev-astro-082214-122302}

\bibitem[{{Ni} {et~al.}(2019){Ni}, {Timlin}, {Brandt}, \& {Yang}}]{Ni+19}
{Ni}, Q., {Timlin}, J., {Brandt}, W.~N., \& {Yang}, G. 2019, Research Notes of the American Astronomical Society, 3, 5, \dodoi{10.3847/2515-5172/aaf8af}

\bibitem[{{Nyland} {et~al.}(2023){Nyland}, {Lacy}, {Brandt}, {Yang}, {Ni}, {Sajina}, {Zou}, \& {Vaccari}}]{nyland+23}
{Nyland}, K., {Lacy}, M., {Brandt}, W.~N., {et~al.} 2023, Research Notes of the American Astronomical Society, 7, 33, \dodoi{10.3847/2515-5172/acbc72}

\bibitem[{{Nyland} {et~al.}(2013){Nyland}, {Alatalo}, {Wrobel}, {Young}, {Morganti}, {Davis}, {de Zeeuw}, {Deustua}, \& {Bureau}}]{nyland+13}
{Nyland}, K., {Alatalo}, K., {Wrobel}, J.~M., {et~al.} 2013, \apj, 779, 173, \dodoi{10.1088/0004-637X/779/2/173}

\bibitem[{{Nyland} {et~al.}(2018){Nyland}, {Harwood}, {Mukherjee}, {Jagannathan}, {Rujopakarn}, {Emonts}, {Alatalo}, {Bicknell}, {Davis}, {Greene}, {Kimball}, {Lacy}, {Lonsdale}, {Lonsdale}, {Maksym}, {Moln{\'a}r}, {Morabito}, {Murphy}, {Patil}, {Prandoni}, {Sargent}, \& {Vlahakis}}]{nyland+18}
{Nyland}, K., {Harwood}, J.~J., {Mukherjee}, D., {et~al.} 2018, \apj, 859, 23, \dodoi{10.3847/1538-4357/aab3d1}

\bibitem[{{Nyland} {et~al.}(2020){Nyland}, {Dong}, {Patil}, {Lacy}, {van Velzen}, {Kimball}, {Sarbadhicary}, {Hallinan}, {Baldassare}, {Clarke}, {Goulding}, {Greene}, {Hughes}, {Kassim}, {Kunert-Bajraszewska}, {Maccarone}, {Mooley}, {Mukherjee}, {Peters}, {Petrov}, {Polisensky}, {Rujopakarn}, {Whittle}, \& {Vaccari}}]{nyland+20}
{Nyland}, K., {Dong}, D.~Z., {Patil}, P., {et~al.} 2020, \apj, 905, 74, \dodoi{10.3847/1538-4357/abc341}

\bibitem[{{O'Dea}(1998)}]{odea+98}
{O'Dea}, C.~P. 1998, \pasp, 110, 493, \dodoi{10.1086/316162}

\bibitem[{{O'Dea} \& {Baum}(1997)}]{odea+97}
{O'Dea}, C.~P., \& {Baum}, S.~A. 1997, \aj, 113, 148, \dodoi{10.1086/118241}

\bibitem[{{O'Dea} \& {Saikia}(2021)}]{odea+21}
{O'Dea}, C.~P., \& {Saikia}, D.~J. 2021, \aapr, 29, 3, \dodoi{10.1007/s00159-021-00131-w}

\bibitem[{{Offringa} {et~al.}(2010){Offringa}, {de Bruyn}, {Biehl}, {Zaroubi}, {Bernardi}, \& {Pandey}}]{offringa+10}
{Offringa}, A.~R., {de Bruyn}, A.~G., {Biehl}, M., {et~al.} 2010, \mnras, 405, 155, \dodoi{10.1111/j.1365-2966.2010.16471.x}

\bibitem[{{Offringa} {et~al.}(2014){Offringa}, {McKinley}, {Hurley-Walker}, {Briggs}, {Wayth}, {Kaplan}, {Bell}, {Feng}, {Neben}, {Hughes}, {Rhee}, {Murphy}, {Bhat}, {Bernardi}, {Bowman}, {Cappallo}, {Corey}, {Deshpande}, {Emrich}, {Ewall-Wice}, {Gaensler}, {Goeke}, {Greenhill}, {Hazelton}, {Hindson}, {Johnston-Hollitt}, {Jacobs}, {Kasper}, {Kratzenberg}, {Lenc}, {Lonsdale}, {Lynch}, {McWhirter}, {Mitchell}, {Morales}, {Morgan}, {Kudryavtseva}, {Oberoi}, {Ord}, {Pindor}, {Procopio}, {Prabu}, {Riding}, {Roshi}, {Shankar}, {Srivani}, {Subrahmanyan}, {Tingay}, {Waterson}, {Webster}, {Whitney}, {Williams}, \& {Williams}}]{offringa+14}
{Offringa}, A.~R., {McKinley}, B., {Hurley-Walker}, N., {et~al.} 2014, \mnras, 444, 606, \dodoi{10.1093/mnras/stu1368}

\bibitem[{{Orienti} \& {Dallacasa}(2008)}]{orienti+08}
{Orienti}, M., \& {Dallacasa}, D. 2008, \aap, 487, 885, \dodoi{10.1051/0004-6361:200809948}

\bibitem[{{Orienti} \& {Dallacasa}(2014)}]{orienti+14}
---. 2014, \mnras, 438, 463, \dodoi{10.1093/mnras/stt2217}

\bibitem[{{Orienti} \& {Dallacasa}(2020)}]{orienti+20}
---. 2020, \mnras, 499, 1340, \dodoi{10.1093/mnras/staa2856}

\bibitem[{{Pasetto} {et~al.}(2016){Pasetto}, {Kraus}, {Mack}, {Bruni}, \& {Carrasco-Gonz{\'a}lez}}]{pasetto+16}
{Pasetto}, A., {Kraus}, A., {Mack}, K.~H., {Bruni}, G., \& {Carrasco-Gonz{\'a}lez}, C. 2016, Astronomische Nachrichten, 337, 91, \dodoi{10.1002/asna.201512271}

\bibitem[{{Patil} {et~al.}(2022){Patil}, {Whittle}, {Nyland}, {Lonsdale}, {Lacy}, {Kimball}, {Lonsdale}, {Peters}, {Clarke}, {Efstathiou}, {Giacintucci}, {Kim}, {Lanz}, {Mukherjee}, \& {Polisensky}}]{patil+22}
{Patil}, P., {Whittle}, M., {Nyland}, K., {et~al.} 2022, \apj, 934, 26, \dodoi{10.3847/1538-4357/ac71b0}

\bibitem[{{Peck} \& {Taylor}(2000)}]{peck+00}
{Peck}, A.~B., \& {Taylor}, G.~B. 2000, \apj, 534, 90, \dodoi{10.1086/308746}

\bibitem[{{Perucho} {et~al.}(2017){Perucho}, {Bosch-Ramon}, \& {Barkov}}]{perucho+17}
{Perucho}, M., {Bosch-Ramon}, V., \& {Barkov}, M.~V. 2017, \aap, 606, A40, \dodoi{10.1051/0004-6361/201630117}

\bibitem[{{Planck Collaboration} {et~al.}(2016){Planck Collaboration}, {Ade}, {Aghanim}, {Arnaud}, {Ashdown}, {Aumont}, {Baccigalupi}, {Banday}, {Barreiro}, {Bartlett}, \& et~al.}]{planck+15}
{Planck Collaboration}, {Ade}, P.~A.~R., {Aghanim}, N., {et~al.} 2016, \aap, 594, A13, \dodoi{10.1051/0004-6361/201525830}

\bibitem[{{Polisensky} {et~al.}(2024){Polisensky}, {Clarke}, {Giacintucci}, \& {Peters}}]{polisensky2024}
{Polisensky}, E., {Clarke}, T.~E., {Giacintucci}, S., \& {Peters}, W. 2024, Frontiers in Astronomy and Space Sciences, 11, 1497375, \dodoi{10.3389/fspas.2024.1497375}

\bibitem[{{Polisensky} {et~al.}(2019){Polisensky}, {Richards}, {Clarke}, {Peters}, \& {Kassim}}]{polisensky+19}
{Polisensky}, E., {Richards}, E., {Clarke}, T., {Peters}, W., \& {Kassim}, N. 2019, in Astronomical Society of the Pacific Conference Series, Vol. 523, Astronomical Data Analysis Software and Systems XXVII, ed. P.~J. {Teuben}, M.~W. {Pound}, B.~A. {Thomas}, \& E.~M. {Warner}, 441

\bibitem[{{Polisensky} {et~al.}(2016){Polisensky}, {Lane}, {Hyman}, {Kassim}, {Giacintucci}, {Clarke}, {Cotton}, {Cleland}, \& {Frail}}]{polisensky+16}
{Polisensky}, E., {Lane}, W.~M., {Hyman}, S.~D., {et~al.} 2016, \apj, 832, 60, \dodoi{10.3847/0004-637X/832/1/60}

\bibitem[{{Punsly} {et~al.}(2021){Punsly}, {Frey}, {Reynolds}, {Marziani}, {Pushkarev}, {Chen}, {Li}, \& {Kharb}}]{punsly+21}
{Punsly}, B., {Frey}, S., {Reynolds}, C., {et~al.} 2021, \apj, 919, 40, \dodoi{10.3847/1538-4357/ac1070}

\bibitem[{{Readhead}(1994)}]{readhead+94}
{Readhead}, A. C.~S. 1994, \apj, 426, 51, \dodoi{10.1086/174038}

\bibitem[{{Readhead} {et~al.}(2024){Readhead}, {Ravi}, {Blandford}, {Sullivan}, {Somalwar}, {Begelman}, {Birkinshaw}, {Liodakis}, {Lister}, {Pearson}, {Taylor}, {Wilkinson}, {Globus}, {Kiehlmann}, {Lawrence}, {Murphy}, {O'Neill}, {Pavlidou}, {Sheldahl}, {Siemiginowska}, \& {Tassis}}]{readhead+24}
{Readhead}, A.~C.~S., {Ravi}, V., {Blandford}, R.~D., {et~al.} 2024, \apj, 961, 242, \dodoi{10.3847/1538-4357/ad0c55}

\bibitem[{{Reines} \& {Volonteri}(2015)}]{reines+15}
{Reines}, A.~E., \& {Volonteri}, M. 2015, \apj, 813, 82, \dodoi{10.1088/0004-637X/813/2/82}

\bibitem[{{Richards} {et~al.}(2011){Richards}, {Max-Moerbeck}, {Pavlidou}, {King}, {Pearson}, {Readhead}, {Reeves}, {Shepherd}, {Stevenson}, {Weintraub}, {Fuhrmann}, {Angelakis}, {Zensus}, {Healey}, {Romani}, {Shaw}, {Grainge}, {Birkinshaw}, {Lancaster}, {Worrall}, {Taylor}, {Cotter}, \& {Bustos}}]{richards+11}
{Richards}, J.~L., {Max-Moerbeck}, W., {Pavlidou}, V., {et~al.} 2011, \apjs, 194, 29, \dodoi{10.1088/0067-0049/194/2/29}

\bibitem[{{Rickett}(1986)}]{rickett+86}
{Rickett}, B.~J. 1986, \apj, 307, 564, \dodoi{10.1086/164444}

\bibitem[{{Rickett}(1990)}]{rickett+90}
---. 1990, \araa, 28, 561, \dodoi{10.1146/annurev.aa.28.090190.003021}

\bibitem[{{Ross} {et~al.}(2021){Ross}, {Callingham}, {Hurley-Walker}, {Seymour}, {Hancock}, {Franzen}, {Morgan}, {White}, {Bell}, \& {Patil}}]{ross+21}
{Ross}, K., {Callingham}, J.~R., {Hurley-Walker}, N., {et~al.} 2021, \mnras, 501, 6139, \dodoi{10.1093/mnras/staa3795}

\bibitem[{{Rujopakarn} {et~al.}(2016){Rujopakarn}, {Dunlop}, {Rieke}, {Ivison}, {Cibinel}, {Nyland}, {Jagannathan}, {Silverman}, {Alexander}, {Biggs}, {Bhatnagar}, {Ballantyne}, {Dickinson}, {Elbaz}, {Geach}, {Hayward}, {Kirkpatrick}, {McLure}, {Micha{\l}owski}, {Miller}, {Narayanan}, {Owen}, {Pannella}, {Papovich}, {Pope}, {Rau}, {Robertson}, {Scott}, {Swinbank}, {van der Werf}, {van Kampen}, {Weiner}, \& {Windhorst}}]{rujopakarn+16}
{Rujopakarn}, W., {Dunlop}, J.~S., {Rieke}, G.~H., {et~al.} 2016, \apj, 833, 12, \dodoi{10.3847/0004-637X/833/1/12}

\bibitem[{{Rujopakarn} {et~al.}(2018){Rujopakarn}, {Nyland}, {Rieke}, {Barro}, {Elbaz}, {Ivison}, {Jagannathan}, {Silverman}, {Smol{\v{c}}i{\'c}}, \& {Wang}}]{rujopakarn+18}
{Rujopakarn}, W., {Nyland}, K., {Rieke}, G.~H., {et~al.} 2018, \apjl, 854, L4, \dodoi{10.3847/2041-8213/aaa9b3}

\bibitem[{{Rybicki} \& {Lightman}(1979)}]{rybicki+79}
{Rybicki}, G.~B., \& {Lightman}, A.~P. 1979, {Radiative processes in astrophysics}

\bibitem[{{Ryu} {et~al.}(2024){Ryu}, {Perna}, \& {Cantiello}}]{ryu+24}
{Ryu}, T., {Perna}, R., \& {Cantiello}, M. 2024, \apjl, 965, L25, \dodoi{10.3847/2041-8213/ad3946}

\bibitem[{{Sheldahl} {et~al.}(2025){Sheldahl}, {Taylor}, {Tremblay}, {Peters}, {Kiehlmann}, {Blandford}, {Lister}, {Pearson}, {Readhead}, {Schinzel}, {Siemiginowska}, \& {Skalidis}}]{sheldahl+25}
{Sheldahl}, E.~E., {Taylor}, G.~B., {Tremblay}, S.~E., {et~al.} 2025, \apj, 987, 26, \dodoi{10.3847/1538-4357/adcc28}

\bibitem[{{Shimwell} {et~al.}(2022){Shimwell}, {Hardcastle}, {Tasse}, {Best}, {R{\"o}ttgering}, {Williams}, {Botteon}, {Drabent}, {Mechev}, {Shulevski}, {van Weeren}, {Bester}, {Br{\"u}ggen}, {Brunetti}, {Callingham}, {Chy{\.z}y}, {Conway}, {Dijkema}, {Duncan}, {de Gasperin}, {Hale}, {Haverkorn}, {Hugo}, {Jackson}, {Mevius}, {Miley}, {Morabito}, {Morganti}, {Offringa}, {Oonk}, {Rafferty}, {Sabater}, {Smith}, {Schwarz}, {Smirnov}, {O'Sullivan}, {Vedantham}, {White}, {Albert}, {Alegre}, {Asabere}, {Bacon}, {Bonafede}, {Bonnassieux}, {Brienza}, {Bilicki}, {Bonato}, {Calistro Rivera}, {Cassano}, {Cochrane}, {Croston}, {Cuciti}, {Dallacasa}, {Danezi}, {Dettmar}, {Di Gennaro}, {Edler}, {En{\ss}lin}, {Emig}, {Franzen}, {Garc{\'\i}a-Vergara}, {Grange}, {G{\"u}rkan}, {Hajduk}, {Heald}, {Heesen}, {Hoang}, {Hoeft}, {Horellou}, {Iacobelli}, {Jamrozy}, {Jeli{\'c}}, {Kondapally}, {Kukreti}, {Kunert-Bajraszewska}, {Magliocchetti}, {Mahatma}, {Ma{\l}ek}, {Mandal}, {Massaro}, {Meyer-Zhao}, {Mingo}, {Mostert}, {Nair},
  {Nakoneczny}, {Nikiel-Wroczy{\'n}ski}, {Orr{\'u}}, {Pajdosz-{\'S}mierciak}, {Pasini}, {Prandoni}, {van Piggelen}, {Rajpurohit}, {Retana-Montenegro}, {Riseley}, {Rowlinson}, {Saxena}, {Schrijvers}, {Sweijen}, {Siewert}, {Timmerman}, {Vaccari}, {Vink}, {West}, {Wo{\l}owska}, {Zhang}, \& {Zheng}}]{shimwell+22}
{Shimwell}, T.~W., {Hardcastle}, M.~J., {Tasse}, C., {et~al.} 2022, \aap, 659, A1, \dodoi{10.1051/0004-6361/202142484}

\bibitem[{{Slob} {et~al.}(2022){Slob}, {Callingham}, {R{\"o}ttgering}, {Williams}, {Duncan}, {de Gasperin}, {Hardcastle}, \& {Miley}}]{slob+22}
{Slob}, M.~M., {Callingham}, J.~R., {R{\"o}ttgering}, H.~J.~A., {et~al.} 2022, \aap, 668, A186, \dodoi{10.1051/0004-6361/202244651}

\bibitem[{{Snellen} {et~al.}(2000){Snellen}, {Schilizzi}, {Miley}, {de Bruyn}, {Bremer}, \& {R{\"o}ttgering}}]{snellen+00}
{Snellen}, I.~A.~G., {Schilizzi}, R.~T., {Miley}, G.~K., {et~al.} 2000, \mnras, 319, 445, \dodoi{10.1046/j.1365-8711.2000.03935.x}

\bibitem[{{Sobolewska} {et~al.}(2019){Sobolewska}, {Siemiginowska}, {Guainazzi}, {Hardcastle}, {Migliori}, {Ostorero}, \& {Stawarz}}]{sobolewska+19}
{Sobolewska}, M., {Siemiginowska}, A., {Guainazzi}, M., {et~al.} 2019, \apj, 871, 71, \dodoi{10.3847/1538-4357/aaee78}

\bibitem[{{Sokolovsky} {et~al.}(2011){Sokolovsky}, {Kovalev}, {Pushkarev}, {Mimica}, \& {Perucho}}]{sokolovsky+11}
{Sokolovsky}, K.~V., {Kovalev}, Y.~Y., {Pushkarev}, A.~B., {Mimica}, P., \& {Perucho}, M. 2011, \aap, 535, A24, \dodoi{10.1051/0004-6361/201015772}

\bibitem[{{Stalevski} {et~al.}(2012){Stalevski}, {Fritz}, {Baes}, {Nakos}, \& {Popovi{\'c}}}]{skirtor2012}
{Stalevski}, M., {Fritz}, J., {Baes}, M., {Nakos}, T., \& {Popovi{\'c}}, L.~{\v{C}}. 2012, \mnras, 420, 2756, \dodoi{10.1111/j.1365-2966.2011.19775.x}

\bibitem[{{Stalevski} {et~al.}(2016){Stalevski}, {Ricci}, {Ueda}, {Lira}, {Fritz}, \& {Baes}}]{skirtor2016}
{Stalevski}, M., {Ricci}, C., {Ueda}, Y., {et~al.} 2016, \mnras, 458, 2288, \dodoi{10.1093/mnras/stw444}

\bibitem[{{Stone} \& {van Velzen}(2016)}]{stone+16}
{Stone}, N.~C., \& {van Velzen}, S. 2016, \apjl, 825, L14, \dodoi{10.3847/2041-8205/825/1/L14}

\bibitem[{{Sullivan} {et~al.}(2024){Sullivan}, {Blandford}, {Begelman}, {Birkinshaw}, \& {Readhead}}]{sullivan+24}
{Sullivan}, A.~G., {Blandford}, R.~D., {Begelman}, M.~C., {Birkinshaw}, M., \& {Readhead}, A. C.~S. 2024, \mnras, 528, 6302, \dodoi{10.1093/mnras/stae322}

\bibitem[{{Sutherland} \& {Bicknell}(2007)}]{sutherland+07}
{Sutherland}, R.~S., \& {Bicknell}, G.~V. 2007, \apss, 311, 293, \dodoi{10.1007/s10509-007-9580-y}

\bibitem[{{Tadhunter}(2016)}]{tadhunter+16}
{Tadhunter}, C. 2016, \aapr, 24, 10, \dodoi{10.1007/s00159-016-0094-x}

\bibitem[{{Tingay} \& {de Kool}(2003)}]{tingay+03}
{Tingay}, S.~J., \& {de Kool}, M. 2003, \aj, 126, 723, \dodoi{10.1086/376600}

\bibitem[{{Tingay} {et~al.}(2015){Tingay}, {Macquart}, {Collier}, {Rees}, {Callingham}, {Stevens}, {Carretti}, {Wayth}, {Wong}, {Trott}, {McKinley}, {Bernardi}, {Bowman}, {Briggs}, {Cappallo}, {Corey}, {Deshpande}, {Emrich}, {Gaensler}, {Goeke}, {Greenhill}, {Hazelton}, {Johnston-Hollitt}, {Kaplan}, {Kasper}, {Kratzenberg}, {Lonsdale}, {Lynch}, {McWhirter}, {Mitchell}, {Morales}, {Morgan}, {Oberoi}, {Ord}, {Prabu}, {Rogers}, {Roshi}, {Udaya Shankar}, {Srivani}, {Subrahmanyan}, {Waterson}, {Webster}, {Whitney}, {Williams}, \& {Williams}}]{tingay+15}
{Tingay}, S.~J., {Macquart}, J.~P., {Collier}, J.~D., {et~al.} 2015, \aj, 149, 74, \dodoi{10.1088/0004-6256/149/2/74}

\bibitem[{{Tremblay} {et~al.}(2009){Tremblay}, {Taylor}, {Helmboldt}, {Fassnacht}, \& {Romani}}]{tremblay+09}
{Tremblay}, S.~E., {Taylor}, G.~B., {Helmboldt}, J.~F., {Fassnacht}, C.~D., \& {Romani}, R.~W. 2009, Astronomische Nachrichten, 330, 206, \dodoi{10.1002/asna.200811157}

\bibitem[{{Tremblay} {et~al.}(2016){Tremblay}, {Taylor}, {Ortiz}, {Tremblay}, {Helmboldt}, \& {Romani}}]{tremblay+16}
{Tremblay}, S.~E., {Taylor}, G.~B., {Ortiz}, A.~A., {et~al.} 2016, \mnras, 459, 820, \dodoi{10.1093/mnras/stw592}

\bibitem[{{Urry} \& {Padovani}(1995)}]{urry+95}
{Urry}, C.~M., \& {Padovani}, P. 1995, \pasp, 107, 803, \dodoi{10.1086/133630}

\bibitem[{{van Bemmel} {et~al.}(2022){van Bemmel}, {Kettenis}, {Small}, {Janssen}, {Moellenbrock}, {Petry}, {Goddi}, {Linford}, {Rygl}, {Liuzzo}, {Marcote}, {Bayandina}, {Schweighart}, {Verkouter}, {Keimpema}, {Szomoru}, \& {van Langevelde}}]{vanbemmel+22}
{van Bemmel}, I.~M., {Kettenis}, M., {Small}, D., {et~al.} 2022, \pasp, 134, 114502, \dodoi{10.1088/1538-3873/ac81ed}

\bibitem[{{van Haarlem} {et~al.}(2013){van Haarlem}, {Wise}, {Gunst}, {Heald}, {McKean}, {Hessels}, {de Bruyn}, {Nijboer}, {Swinbank}, {Fallows}, {Brentjens}, {Nelles}, {Beck}, {Falcke}, {Fender}, {H{\"o}randel}, {Koopmans}, {Mann}, {Miley}, {R{\"o}ttgering}, {Stappers}, {Wijers}, {Zaroubi}, {van den Akker}, {Alexov}, {Anderson}, {Anderson}, {van Ardenne}, {Arts}, {Asgekar}, {Avruch}, {Batejat}, {B{\"a}hren}, {Bell}, {Bell}, {van Bemmel}, {Bennema}, {Bentum}, {Bernardi}, {Best}, {B{\^\i}rzan}, {Bonafede}, {Boonstra}, {Braun}, {Bregman}, {Breitling}, {van de Brink}, {Broderick}, {Broekema}, {Brouw}, {Br{\"u}ggen}, {Butcher}, {van Cappellen}, {Ciardi}, {Coenen}, {Conway}, {Coolen}, {Corstanje}, {Damstra}, {Davies}, {Deller}, {Dettmar}, {van Diepen}, {Dijkstra}, {Donker}, {Doorduin}, {Dromer}, {Drost}, {van Duin}, {Eisl{\"o}ffel}, {van Enst}, {Ferrari}, {Frieswijk}, {Gankema}, {Garrett}, {de Gasperin}, {Gerbers}, {de Geus}, {Grie{\ss}meier}, {Grit}, {Gruppen}, {Hamaker}, {Hassall}, {Hoeft}, {Holties},
  {Horneffer}, {van der Horst}, {van Houwelingen}, {Huijgen}, {Iacobelli}, {Intema}, {Jackson}, {Jelic}, {de Jong}, {Juette}, {Kant}, {Karastergiou}, {Koers}, {Kollen}, {Kondratiev}, {Kooistra}, {Koopman}, {Koster}, {Kuniyoshi}, {Kramer}, {Kuper}, {Lambropoulos}, {Law}, {van Leeuwen}, {Lemaitre}, {Loose}, {Maat}, {Macario}, {Markoff}, {Masters}, {McFadden}, {McKay-Bukowski}, {Meijering}, {Meulman}, {Mevius}, {Middelberg}, {Millenaar}, {Miller-Jones}, {Mohan}, {Mol}, {Morawietz}, {Morganti}, {Mulcahy}, {Mulder}, {Munk}, {Nieuwenhuis}, {van Nieuwpoort}, {Noordam}, {Norden}, {Noutsos}, {Offringa}, {Olofsson}, {Omar}, {Orr{\'u}}, {Overeem}, {Paas}, {Pandey-Pommier}, {Pandey}, {Pizzo}, {Polatidis}, {Rafferty}, {Rawlings}, {Reich}, {de Reijer}, {Reitsma}, {Renting}, {Riemers}, {Rol}, {Romein}, {Roosjen}, {Ruiter}, {Scaife}, {van der Schaaf}, {Scheers}, {Schellart}, {Schoenmakers}, {Schoonderbeek}, {Serylak}, {Shulevski}, {Sluman}, {Smirnov}, {Sobey}, {Spreeuw}, {Steinmetz}, {Sterks}, {Stiepel}, {Stuurwold},
  {Tagger}, {Tang}, {Tasse}, {Thomas}, {Thoudam}, {Toribio}, {van der Tol}, {Usov}, {van Veelen}, {van der Veen}, {ter Veen}, {Verbiest}, {Vermeulen}, {Vermaas}, {Vocks}, {Vogt}, {de Vos}, {van der Wal}, {van Weeren}, {Weggemans}, {Weltevrede}, {White}, {Wijnholds}, {Wilhelmsson}, {Wucknitz}, {Yatawatta}, {Zarka}, \& {Zensus}}]{van_haarlem+13}
{van Haarlem}, M.~P., {Wise}, M.~W., {Gunst}, A.~W., {et~al.} 2013, \aap, 556, A2, \dodoi{10.1051/0004-6361/201220873}

\bibitem[{{Vermeulen} {et~al.}(2003){Vermeulen}, {Ros}, {Kellermann}, {Cohen}, {Zensus}, \& {van Langevelde}}]{vermeulen+03}
{Vermeulen}, R.~C., {Ros}, E., {Kellermann}, K.~I., {et~al.} 2003, \aap, 401, 113, \dodoi{10.1051/0004-6361:20021752}

\bibitem[{{Wagner} \& {Witzel}(1995)}]{wagner+95}
{Wagner}, S.~J., \& {Witzel}, A. 1995, \araa, 33, 163, \dodoi{10.1146/annurev.aa.33.090195.001115}

\bibitem[{{Walker}(1998)}]{walker+98}
{Walker}, M.~A. 1998, \mnras, 294, 307, \dodoi{10.1046/j.1365-8711.1998.01238.x10.1111/j.1365-8711.1998.01238.x}

\bibitem[{{We{\.z}gowiec} {et~al.}(2024){We{\.z}gowiec}, {Jamrozy}, {Chy{\.z}y}, {Hardcastle}, {Ku{\'z}micz}, {Heald}, \& {Shimwell}}]{wezgowiec+24}
{We{\.z}gowiec}, M., {Jamrozy}, M., {Chy{\.z}y}, K.~T., {et~al.} 2024, \aap, 691, A193, \dodoi{10.1051/0004-6361/202451580}

\bibitem[{{York} {et~al.}(2000){York}, {Adelman}, {Anderson}, {Anderson}, {Annis}, {Bahcall}, {Bakken}, {Barkhouser}, \& {et al.}}]{york+00}
{York}, D.~G., {Adelman}, J., {Anderson}, Jr., J.~E., {et~al.} 2000, \aj, 120, 1579, \dodoi{10.1086/301513}

\bibitem[{{Zinn} {et~al.}(2012){Zinn}, {Middelberg}, {Norris}, {Hales}, {Mao}, \& {Randall}}]{zinn+12}
{Zinn}, P.~C., {Middelberg}, E., {Norris}, R.~P., {et~al.} 2012, \aap, 544, A38, \dodoi{10.1051/0004-6361/201219349}

\bibitem[{{Zovaro} {et~al.}(2019){Zovaro}, {Sharp}, {Nesvadba}, {Bicknell}, {Mukherjee}, {Wagner}, {Groves}, \& {Krishna}}]{zovaro+19}
{Zovaro}, H. R.~M., {Sharp}, R., {Nesvadba}, N. P.~H., {et~al.} 2019, \mnras, 484, 3393, \dodoi{10.1093/mnras/stz233}

\end{thebibliography}
\bibliographystyle{aasjournal}


\listofchanges

\end{document}